\documentclass[preprint]{elsarticle}
\usepackage{multirow, booktabs}
\usepackage{colortbl}
\usepackage{tikz}
\usepackage{amsmath, amssymb}
\usepackage{algorithm,algpseudocode}
\usepackage{listings}
\usepackage{pgfplots}
\usepackage[bf,BF]{subfigure}
\usepackage{hyperref}

\usepackage{siunitx}

\usepackage{graphicx}
\usepackage{lscape}
\usepackage{array}
\usepackage{comment}
\newcolumntype{P}[1]{>{\centering\arraybackslash}p{#1}}

\usepackage[normalem]{ulem}
\usepackage{color}

\newcommand{\rev}[1]{#1}
\newcommand{\revd}[1]{}
\newcommand{\revdd}[2]{#2}

\AtBeginDocument{%
  \providecommand\BibTeX{{%
    \normalfont B\kern-0.5em{\scshape i\kern-0.25em b}\kern-0.8em\TeX}}}

\begin{document}
\begin{frontmatter}

\title{Effects of lower floating-point precision on scale-resolving numerical simulations of turbulence}

\author[1]{Martin Karp}
\ead{makarp@kth.se}
\author[1]{Ronith Stanly}
\author[2]{Timofey Mukha}
% \ead{timofey.mukha@kaust.edu.sa}
\author[2,3]{Luca Galimberti}
% \ead{luca.galimberti@kaust.edu.sa}
\author[4]{Siavash Toosi}
\author[5]{Hang Song}
\author[2]{Lisandro Dalcin}
% \ead{lisandro.dalcin@kaust.edu.sa}
\author[6]{Saleh Rezaeiravesh}
\author[1]{Niclas Jansson}
% \ead{njansson@kth.se}
\author[1]{Stefano Markidis}
\author[2]{Matteo Parsani}
% \ead{matteo.parsani@kaust.edu.sa}
\author[5]{Sanjeeb Bose}
\author[5]{Sanjiva Lele}
\author[1,4]{Philipp Schlatter}

\address[1]{KTH Royal Institute of Technology, Stockholm, Sweden}
\address[2]{King Abdullah University of Science and Technology, Thuwal, Saudi Arabia}
\address[3]{Politecnico di Milano, Milan, Italy}
\address[4]{Friedrich--Alexander--Universität (FAU) Erlangen-Nürnberg, Erlangen, Germany}
\address[5]{Stanford University, Stanford, USA}
\address[6]{The University of Manchester, Manchester, UK}

\begin{abstract}
Modern computing clusters offer specialized hardware for reduced-precision arithmetic, which can significantly speed up the time to solution. This is possible due to a decrease in data movement, as well as the ability to perform arithmetic operations at a faster rate. However, for high-fidelity simulations of turbulence, such as direct and large-eddy simulation, the impact of reduced precision on the computed solution and the resulting uncertainty across flow solvers and different flow cases has not been explored in detail\rev{,} and limits the optimal utilization of new high-performance computing systems. 
%Hence, a better understanding of how reduced-precision affects large-scale CFD simulations is essential.
In this work, the effect of reduced precision is studied using four diverse computational fluid dynamics (CFD) solvers (two incompressible, Neko and Simson, and two compressible, PadeLibs and SSDC) using four test cases: turbulent channel flow at $Re_{\tau}=550$ and higher, forced transition in a channel, flow over a cylinder at $Re_{D}=3900$, and compressible flow over a wing section at $Re_c=50000$. We observe that the flow physics are remarkably robust with respect to reductions in lower floating-point precision, and that often other forms of uncertainty, due to, for example, time averaging, often have a much larger impact on the computed result. Our results indicate that different terms in the Navier--Stokes equations can be computed to a lower floating-point accuracy without affecting the results. In particular, standard IEEE single precision can be used effectively for the entirety of the simulation, showing no significant discrepancies from double-precision results across the solvers and cases considered. Potential pitfalls are also discussed.
\end{abstract}

\begin{keyword}
Computational Fluid Dynamics, turbulence, direct numerical simulation, floating-point precision
\end{keyword}
\end{frontmatter}

\section{Introduction} \label{sec:introduction}
Computational fluid dynamics (CFD) has become an essential tool in both academic research and industry, encompassing a wide range of applications. Over time, a variety of models and numerical methods have been developed to integrate the governing equations of fluid motion, each tailored to specific use cases, desired accuracy, and computational constraints.
Yet, until relatively recently, all of them typically relied on IEEE double precision floating point numbers (FP64) to numerically compute the solution. 

\begin{figure}
    \centering
    \includegraphics[width=\linewidth]{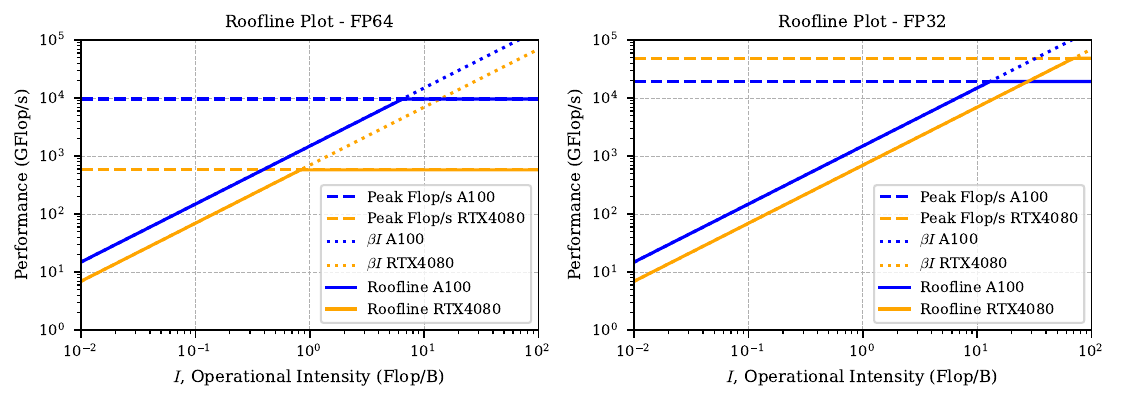}
    \caption{Roofline for the Nvidia A100 and Nvidia GeForce RTX4080 in double (FP64) and single (FP32) precision. The solid lines represent the roofline (the maximum attainable performance) for the two architectures as a function of operational intensity $I=\pi/\beta$, defined as the fraction between the peak performance $\pi$ and the memory bandwidth $\beta$ of the computing unit. The dashed lines represent the peak performance $\pi$ for the two architectures and the dotted line the performance limit based on the time needed to load data from memory, $\beta I$. Most CFD codes today operate in the domain limited by $\beta I$.}
    \label{fig:roofline}
\end{figure}

Due to recent shifts in hardware manufacturing tailored to decrease the energy consumption of floating-point operations and drive up performance for artificial intelligence (AI) applications, hardware support for lower precision floating-point numbers has become increasingly prevalent~\cite{dongarra2024hardware}. 
On these new platforms, lower precision offers both higher performance and higher energy efficiency, as well as a smaller memory footprint, reducing the amount of data movement necessary. All of these developments ultimately lead to significant savings in both time and energy, and thus provide considerable monetary savings for large-scale computations.
As CFD practitioners, where most codes are limited by memory bandwidth due to large relatively sparse linear algebra systems, the use of FP32 instead of FP64 would, for example, move the bandwidth roofline by a factor of 2, indicating an up to 2X of performance would be attainable for a bandwidth-limited code~\cite{williams2009roofline}. A simple roofline comparison between two relatively common GPUs, Nvidia A100 and Nvidia GeForce RTX4080, is shown in Figure~\ref{fig:roofline}. 
The simple performance model clearly illustrates how the roofline for the RTX4080, with a significantly lower performance for FP64, improves for FP32 and is comparable to the more expensive A100. Considering that the operational intensity $I$ for a given code also improves with lower precision, this suggests that significant gains can be enabled by lowering precision.
In practice, a factor of $2\times$ in performance is rarely achieved due to other factors such as kernel launch latencies or communication overhead in parallel communications.

% However, 
While there are clear opportunities for performance gains, an important question is to what extent these lower floating point formats can be used for scale-resolving simulations of turbulence without sacrificing the required accuracy. This is what we aim to assess in this work.

This question is currently under active investigation by the research community and can be viewed as part of a broader trend across multiple disciplines~\cite{kashi2024mixed}.
A full review of works on lower-precision arithmetics and CFD is outside the scope of this article. Still, a few selected references are provided for the benefit of the reader. 
The first example is from simulations of weather and climate. Several articles from Klöwer, Paxton, Düben, and Palmer et al.~\citep{klower2020number,klower2023periodic,paxton2022climate} highlight that lower precision and less exact computing open up for large performance improvements.
In the realm of compressible flows, \revdd{the CFD solver PyFR, \mbox{\citep{witherden2020impact,pyfr-VERMEIRE2017-jcp}},}{Witherden and Jameson\mbox{~\citep{witherden2020impact}} and Vermeire et al.\mbox{~\citep{pyfr-VERMEIRE2017-jcp}}} used single precision (FP32) in the context of a high-order flux-reconstruction scheme (discontinuous Galerkin) for implicit large eddy simulation for a wide range of cases using the CFD solver PyFR. 
Methods that enable mixed-precision computing while maintaining acceptable levels of accuracy have been proposed and investigated, for instance, in the context of the finite volume solver OpenFOAM~\citep{mixed_prec_openfoam-Delorme2025}. Further encouraging results using OpenFOAM were also found by Brogi et al.~\citep{brogi2024floating}. 
Haridas et al.~\citep{neural_network-HARIDAS2022} explored the possibility of using neural networks to correct errors introduced using reduced precision arithmetic in simulating fluid dynamic problems. 
The feasibility of conducting mixed-precision operations in the context of high-order compact finite difference schemes was investigated by Song et al.~\cite{song2023mixed} using the PadeLibs code for CFD. Wang et al.~\citep{WANG2025} proposed a mixed precision strategy in the finite volume method for unstructured grids that used high precision near solid bodies and lower precision far away from them. Freytag et al. \cite{gabriel-2022} studied the performance and power efficiency of using reduced- and mixed-precision arithmetic for CFD. Bhola and Duraisamy~\cite{bhola2024deterministic,bhola2024exploiting} performed analys\rev{e}s of errors incurred due to rounding in mixed precision computations. Walden et al. \cite{nasa_fun3d_redprec2019} investigated the speedup achieved on GPUs using the FUN3D CFD code with reduced-precision arithmetic for their memory-bound linear solver kernel.
Grout \cite{spectral_deffered} used reduced-precision for the low-order time integration methods employed to construct higher-order methods in the so-called spectral differed correction method used in combustion CFD and studied the resulting rate of convergence. 
To reduce the communication bottleneck, 
\rev{Reuther et al.~\cite{reuther:97} in the context of a multiblock RANS solver and}
Rogowski et al.~\cite{matteo2021}
\rev{in the context of discontinuous Galerkin (using the SSDC solver), among others,} 
investigated \rev{the impact and performance of} reduced precision for \rev{parallel} communications.
\revd{in the context of Discontinuous Galerkin method using the SSDC CFD code.}
\rev{More recently, Siklòsi et al.~\cite{siklosi2026reducedprecision} explored the use of mixed precision for compressible turbulent flow computations with an explicit finite difference framework.}
Finally, for CFD based on the Lattice Boltzmann method rather than solving the Navier-Stokes (NS) equations,~\cite{lehmann2022accuracy} demonstrates the possibility to accelerate the solver using lower precision.
It is also worth noting that finite volume codes used in industry, both open-source and commercial, offered the possibility to compile in FP32 long before the recent increase in low-precision computer hardware. There are also many mixed-precision algorithms for important linear algebra operations used in CFD, such as preconditioners and iterative solvers, to obtain a result in full double precision~\citep{fischer2022nekrs,higham2022mixed,kashi2024mixed,abdelfattah2021survey}.
This potentially suggests a wider scale of adoption of FP32 than what is documented in the scientific literature.

Despite these success stories, it is important to acknowledge that, due to the nonlinear nature of the NS equations, cases can be found where numerical precision has a larger-than-expected impact.
In particular, in the context of dynamical systems, if there exist several attractors of the flow or a symmetry that is sensitive to small disturbances, lower floating-point precision can be detrimental to the validity of the simulation \citep{fleischmann2019numerical,qin2022large}. However, if the simulation only has one attractor\revdd{,}{} and the results are insensitive to small thermal fluctuations, which, in many ways\rev{,} can\revd{,} act similarly to low-precision~\cite{liao2025physical}, lower precision might be utilized effectively.

The question in the focus of this article is whether reduced precision arithmetic can be successfully employed in direct numerical simulations (DNS) and wall-resolved large-eddy simulation (LES).
The authors' impression is that the DNS community is generally skeptical about abandoning FP64.
There are good reasons behind that.
DNS are usually conducted to obtain the ground truth for a given flow and study all possible subtleties in its behavior.
%Examples include analysis of coherent structures, spectra, and higher-order statistics.
Therefore, even small errors are considered unacceptable.
Furthermore, a new DNS is usually conducted at the limit of the computing budget, and simply trying out a low-precision simulation can be perceived as not worth the risk.

Nevertheless, the main outcome of the DNS is often primarily expressed in a statistical description of the flow, and an argument can be made that it must, to a degree, be robust to small numerical errors, the origin of which may be both discretization and the precision of arithmetics.
Indeed, there are already documented successful attempts in using FP32 for DNS of homogeneous isotropic turbulence (HIT)~\citep{homann2007impact, yeung2015extreme, yeung2018effects, yeung2020advancing}, with no discernible differences observed between the FP32 and FP64 simulations. 
While this is very encouraging, to trigger a shift across the wider high-fidelity CFD community, it is necessary for similar evidence to emerge across a broader range of flow cases and numerical techniques.
The goal of the current work is to be a step in that direction\revdd{,}{} and toward a deeper understanding of the potential impact of lower floating precision on high-fidelity CFD. 
This builds on some preliminary work performed by Karp et al.~\cite{karp2023,karp2024-ctr,karp2025robustness}\rev{, and should be viewed as an extension of the existing evidence in the literature by investigating
a wider range of solvers, flows, and arithmetic precisions}.

To that end, our study employs four different solvers (Neko, SSDC, Simson, PadeLibs) which differ in both the formulation of the governing equations and the approach to discretization.
We consider four use cases, with a focus on wall-bounded flows mainly in the turbulent regime.
The latter is motivated, in part, by the scientific interests of the authors, but also by the fact that turbulence near the wall has a particularly rich and complex structure that could potentially be disrupted due to floating\rev{-}point errors.
The four use cases cover several important flow classes: internal flows (the channel flow test case), transition to turbulence (the Tollmien-Schlichting wave test case), external flows over bluff bodies (flow around a cylinder), and external flows with influence of compressibility (the NACA-0012 case).

To extend the range of arithmetic precisions, we go below FP32 and also consider precision that is not necessarily implemented in current hardware.
To facilitate that, we rely on a software emulation of the precision.
However, we have also developed native FP32 versions of Neko, Simson, and, partially, SSDC.
This work reveals that naively reducing the precision of all reals may lead to various pitfalls.
Therefore, in addition to presenting simulation results, we also summarize our practical experiences in adapting our codes to FP32, in order to aid other members of the community in similar efforts.
\rev{While not used in the work, tools such as the Herbie project~\citep{herbie}, which aim at identifying floating\rev{-}point problems in the code, may be of particular interest when adapting the codes for lower precisions.}

The paper is structured as follows. Section 2 provides a description of the common floating-point formats. Section 3 introduces the four CFD codes that will be used for the present study. The remainder of the paper presents the impact of precision on four different flow cases: Turbulent channel flow will be analysed in Section 4, and Section 5 treats transitional channel flow. External flow with separation is discussed in Section 6, followed by the flow around wings in Section 7. Practical experiences and conclusions wrap up the paper in Sections 8 and 9.

\section{Floating-point numbers} \label{sec:numbers}
In this work, we are interested in the impact of floating\rev{-}point formats and their ability to represent relevant physics in a turbulent flow simulation. A floating-point number is defined by a number of mantissa bits $b$ and exponent bits $b_e$ together with one sign bit $s$ dictating the sign of the floating-point number. If we let $e$ be the value of the exponent (as an unsigned $b_e$-bit integer), and $c_i$ be the $i$th least significant bit of the mantissa, the value for a given normal floating-point number is 
\begin{equation}
(-1)^s \left(1 + \sum_{i=1}^{b} c_{b - i} 2^{-i}\right ) \times 2^{e - (2^{b_e-1}-1)}.
\end{equation}
We consider floating-point formats ranging from 8  to 64 bits, as listed in Table~\ref{tab:FP}. In addition to the normal floating-point numbers, there are also special numbers such as $\pm$ infinity, not a number (NaN), and subnormal numbers.

\begin{table}
\centering
    \begin{tabular}{lllll }
        \hline
        Name & bits& $b$ & $b_e$ & {\centering $\varepsilon=2^{-b-1}$}  \\
        \hline
        \textbf{FP64}\rule{0pt}{2.4ex}& 64 & 52 & 11&$2^{-53}\approx 2\cdot 10^{-16}$ \\
        \textbf{FP32} & 32 & 23 & 8&$2^{-24}\approx 6\cdot 10^{-8}$  \\
        \textbf{FP16} & 16 & 10 & 5&$2^{-11}\approx 5\cdot 10^{-4}$ \\        
        \textbf{bfloat16} & 16 & 7 & 8 & $2^{-8}\approx 4\cdot 10^{-3}$  \\
        \textbf{E4M3} & 8 & 3 &4 &$2^{-4} = 0.0625$  \\
        \textbf{E5M2} & 8 & 2 & 5&$2^{-3} = 0.125$  \\
        \hline
    \end{tabular}
    \caption{Different floating-point formats.
    % ranging from FP64 to the recent FP8 formats E4M3 and E5M2. 
    The rounding machine epsilon $\varepsilon$,  $|u_{\text{FP}} - u| < \varepsilon u$ for some real number $u$, is the largest round-off error introduced due to floating-point precision.}
    \label{tab:FP}   
\end{table}

Floating-point numbers have a great strength in their large range and the use of a relative rounding error, compared to an absolute one, such as in fixed-point formats. This means that when operations are carried out on numbers of comparable amplitude, the effect of rounding error is relatively small. However, issues can arise, for example, when performing computations with two numbers $x$ and $y$ where $x \ll y$. In this case, if $x < \varepsilon y$, the computation may be subject to so-called \textit{stagnation}. A typical example is when performing summations of long arrays, and an individual element in the array is smaller than the total sum, which can yield a significant rounding error. This would, for example, yield inaccurate dot products, large errors after many time steps, as well as possibly impact the collection of statistical quantities.

This aspect of stagnation is a general issue in dynamical systems as well, and efforts to avoid it through approaches such as stochastic rounding have been suggested \cite{paxton2022climate}. However, floating-point numbers with deterministic rounding (rounding to the nearest) are the most readily available and commonly used in modern computing systems, and in this work we limit ourselves to this type of quantization. 

The issue of when the machine epsilon plays a large role for CFD, can be clearly seen through inspection of the NS equations as well. If we consider the non-dimensionalized incompressible NS equations, 

\begin{equation}\label{eq:incom}
\begin{split}
    \nabla  \cdot  \mathbf{v} &= 0, \\ 
    \frac{\partial \mathbf{v}}{\partial t} + (\mathbf{v}\cdot \nabla) \mathbf{v} &= - \nabla p + \frac{1}{Re}\nabla^2 \mathbf{v} + \mathbf{F},
\end{split}
\end{equation}
where $\mathbf{v}$ is the velocity field, $p$ the pressure, $\mathbf{F}$ an external forcing and $Re = U/\nu L$ is the Reynolds number defined for some suitable characteristic velocity $U$ and length scale $L$ and the kinematic viscosity of the fluid $\nu$. As can be seen, the choice of non-dimensionalization and the $Re$ number immediately provides a connection between the strength of the different terms of the equations. Taking the extreme case when for example $\frac{1}{Re} < \varepsilon$, the time integration of the system would be significantly affected by numerical round-off.

If we consider $\mathbf{u}^i$ to be the state of the flow at time step $i$ and time $i\Delta t$, it is a discretized vector of length $n$, $\mathbf{u} \in FP^{n}$, where $FP$ is the set of floating point numbers possible to be represented for a floating-point format FP. The total simulation up to time $i\Delta t $ can then be  described as the set of realizations of the flow $\mathbf{U} = \{u^0,\ldots,u^i\}$, where $\mathbf{U}$ is constructed through some map from state $\mathbf{u}^i$ to $\mathbf{u}^{i+1}$
\begin{equation}\label{eq:state}
    \mathbf{u}^{i+1} = f(\mathbf{u}^{i}).
\end{equation}

In this investigation we make an attempt to see \rev{the impact of floating point precision on different terms in the governing equations and how this affects the final discretized system $f$ and its behavior across numerical methods irrespective of their state representation $\mathbf{u}$.} \revd{ how $f$, which is described by the numerical method, is affected, and the elements in $\mathbf{u}$ are not the focus, but rather how different terms in the Navier-Stokes equations and the state vector $\mathbf{u}$ are affected by lower floating point precision.} This view of a numerical simulation holds for any numerical discretization of the flow. Evaluating different numerical methods, such as finite volume, finite elements, or similar, would in this model equal the choice of the element in the state vector and how the map $f$ is computed. 

\section{Methodology} \label{sec:methodology}
Throughout this work, we consider floating-point precision as a rounding operation from some state $u$ to a rounded state $\tilde u$.  In our work, for all solvers considered in the following sections, all floating-point numbers below FP32 are emulated with CPFloat~\citep{fasi2023cpfloat}.
Several opportunities exist to incorporate rounding into the governing equations, and the question is how different terms, such as the advective nonlinear term, depend on the numerical precision. The nonlinear term is especially important for turbulent cases where $Re$ is large, the solution to the  equations is chaotic, and small disturbances cause two trajectories to diverge quickly. As such, analytical or deterministic approaches to assess the accuracy of the simulation are no longer applicable.  In addition, in a numerical setting where the equations are discretized, the interaction and dependence on an accurate numerical format can be drastic. As such, we consider several different cases to evaluate whether a simulation can be run entirely in lower precision, as well as whether only specific terms in the governing equations need to be perturbed. In particular, we distinguish three approaches:

\begin{enumerate}
    \item \textbf{Full FP32.} The entire solver is run using IEEE single precision. This is the only case where the lower precision is not emulated in this work.
    \item \textbf{State rounding.} Casting $\boldsymbol{u}^i$ in lower precision, while the solver operates in FP64. Simulating the perturbed system, $ \mathbf{u}^{i+1} = f(\tilde{\mathbf{u}^{i}})$, where the state is constrained to a lower floating point precision at each time $i$.
    \item \textbf{Term rounding.} Different terms in the NS equations, such as the convective or viscous term or both, are represented in lower precision; for example, the convective term in the incompressible formulation would be computed according to $\widetilde{(\mathbf{v}\cdot \nabla) \mathbf{v}}$ with the rounding operator. 
\end{enumerate}  

To assess the impact of these perturbations on various numerical schemes and formulations of the incompressible and compressible Navier--Stokes equations, we consider four different flow solvers with distinct discretizations and characteristics.
In light of these differences, the specific way the rounding is applied differs between the codes.
These details are discussed for each solver individually.

\subsection{Software and numerical methods}
\subsubsection{Neko}
Neko is based on a continuous Galerkin spectral-element framework with a special focus on the incompressible Navier--Stokes equations, with extensive support for heterogeneous computer architectures~\citep{jansson2024neko}. The code has excellent scaling demonstrated up to thousands of GPUs and was nominated for the Gordon Bell Prize in 2023~\citep{neko_gordon_bell2023}. The solver uses high-order hexahedral spectral elements (polynomial order 7 for the tests here), with the $P_N-P_N$ method for velocity--pressure decoupling, a third-order semi-implicit time integration method, and dealiasing of the convective term using the $3/2$-rule~\citep{deville2002high,nek5000_2021_tech_report}. 
The following tests are performed with Neko: Full FP32 (representing all floating-point numbers and executing operations in FP32), perturbation of the convective term (denoted Convective/Conv. FPX in tables and plots for precision FPX), and state rounding (denoted State FPX). Note that for the Full FP32 simulations, the mesh files used in the simulations were still the same, but were directly converted to FP32 when loaded into Neko.

\subsubsection{Simson}\label{sec:simsonintro}
Simson~\cite{chevalier_schlatter_lundbladh_henningson_2007} is a fully spectral code for channel and boundary-layer configurations, based on Fourier discretization in the streamwise and spanwise directions, and Chebyshev expansion in the vertical (wall-normal) direction.
The mesh is equidistant in the wall-parallel directions, and follows a Gauss--Lobatto--Chebyshev distribution in the wall-normal direction. Standard dealiasing using the $3/2$ rule is performed in the Fourier directions only. All solvers are direct in velocity--vorticity formulation; thus, no tolerances need to be specified. 
The classical fourth-order Runge-Kutta (RK4) method is used for time integration. 
Tests performed using Simson include Full FP32, Convective FPX, and State FPX.
Convective FPX is implemented in Simson by rounding the convective term after its calculation.
State rounding is implemented by rounding of all relevant fields (velocities and vorticities) at the start of each RK4 substep. 
Since precisions lower than FP32 are emulated,
in both State and Convective FPX
the rest of the operations are performed in FP64.

\subsubsection{SSDC}\label{sec:ssdcintro}
SSDC implements a high-order entropy-stable discontinuous collocated Galerkin method for the compressible Navier--Stokes equations~\cite{parsani2021ssdc} with support for heterogeneous computer architectures.
The SSDC framework is built on top of the highly scalable Portable and Extensible Toolkit for Scientific Computing (PETSc) \cite{petsc-user-ref}, its mesh topology abstraction (DMPlex), and its scalable differential–algebraic equation solver components. The spatial discretization features $hp$-adaptive capabilities on unstructured quadrilateral/hexahedral meshes. Support for nonconforming meshes relies on the p4est software library \cite{BursteddeWilcoxGhattas11} and its bridge to PETSc's DMPlex. Triangle/tetrahedral meshes are converted on the fly into quadrilateral/hexahedral elements; uniform and non-uniform mesh refinement algorithms are also available. The collocation nodes inside each element are distributed according to the 
Gauss--Lobatto--Legendre 
quadrature points. The solver has demonstrated excellent  strong parallel scaling up to 864\,000 CPU cores and thousands of GPUs on the Shaheen III supercomputer hosted at KAUST. The time integration is explicit and performed using the Runge--Kutta scheme of Bogacki--Shampine \cite{bogacki1989rungekutta} of order three with four stages with the first-same-as-last property. This Runge--Kutta scheme has an embedded second-order method used to implement adaptive step size. 

Full FP32 capabilities were partially implemented in SSDC as part of this work, currently limited to precomputed metric terms in FP64 on top of a fully FP32 computation (see Section~\ref{sec:practical_experiences} for details).
State rounding was not considered.
Regarding term-rounding, the explicit time integration implies that the rounding is applied to the corresponding term in the right-hand side, at each stage of the Runge--Kutta scheme.
We implemented the possibility to treat the viscous and convective operators separately.
Moreover, the rounding can be applied either to the fields before evaluating the operator, to the evaluated operator's result, or to both.
To be aligned with the notation for the other codes, Convective FPX refers to rounding the convective operator's output.
Applying rounding to both terms, prior to and after the application of the respective operators is referred to as Combined FPX.

\subsubsection{PadeLibs}
PadeLibs is a Navier--Stokes solver for high-resolution simulations of compressible turbulent flows \cite{song2024robust}. The numerical discretization uses sixth-order compact finite-difference methods 
with collocated variable storage and staggered flux assembly. The simulation framework used in PadeLibs is robust to aliasing errors and has high accuracy in resolving diffusive fluxes at small scales. 
In this work, round-off effects are investigated by rounding the convective (inviscid) fluxes to a precision FPX (Convective FPX) after they are assembled before taking the divergence operations. 
The rounded results still keep the double-precision format (FP64), although the emulated round-off errors are introduced. All the differential and interpolation operations are consistently calculated in double-precision format. The operator coefficients are all at double-precision accuracy, and the round-off errors are added only from the input. For the incompressible test cases, the Mach number is set to be $0.25$.

%%%%%%%%%%%%%%%%%%%%%%%%%%%%%%%%%%%%%%%%%%%%%%%%%%%%%%%%%%%%%%%%%%%%%
 
\section{Fully developed turbulence} \label{sec:channel}

The first test case is turbulent channel flow at $Re_{\tau}=550$ in a relatively modest domain of $2\pi\delta \times 2\delta \times \pi \delta$.
Here, $\delta$ is the half-height of the channel and $Re_\tau$ is the friction Reynolds number.
The resolutions follow standard practice for high-order simulations of wall turbulence: $\Delta x^+ \approx 12$, $\Delta z^+ \approx 5$, and $\Delta y^+$ similar to, e.g.~,~\cite{delalamo}.
In particular, for SSDC and Neko, the first off-wall node is located at $y^+ \approx 0.45$.
Table \ref{tab:channel} summarizes the different successful simulations and their corresponding simulation parameters.
The maximum difference of the first- and second-order moments for the streamwise velocity, $u$, in inner units is compared with that of the FP64 simulation 
using the respective code as well as with the DNS of Lee and Moser (LM) \cite{Lee_Moser_2015}. These error measures are denoted using the expression shown in Equation~(\ref{eq:error_channel}), where $q$ can be $u$ or $u'u'$, and $\rm ``ref"$ is replaced either with ``FP64" if the FP64 simulation
(from the corresponding CFD code) is used, or is replaced with ``LM" if the data from LM~\cite{Lee_Moser_2015} is used instead. 

\begin{equation}
    \mathcal{E}_{q_{\rm ref}} = \text{max}\left(|\langle q\rangle^{+}-\langle q\rangle^{+}_{\rm ref}|\right)
    \label{eq:error_channel}
\end{equation}

\begin{table}
    \centering
\resizebox{\textwidth}{!}{%

    \begin{tabular}{llllcccc}
         Setup& $Re_\tau$ & Avg. time & $\mathcal{E}_{U_{\rm FP64}}$ & $\mathcal{E}_{u'u'_{\rm FP64}}$ & $\mathcal{E}_{U_{\rm LM}}$ & $\mathcal{E}_{u'u'_{\rm LM}}$\\ \\
         \hline
         Neko &&&&&&&\\
     Full FP64 & 548 &$66.6\delta/u_\tau$ & --- & --- & 0.008 & 0.125\\
     Full FP32 & 557 &$150.0\delta/u_\tau$ & 0.179 & 0.126 &  0.154 & 0.107\\
     Convective FP32 & 550 &$105.5\delta/u_\tau$ & 0.109 & 0.103 & 0.090 & 0.112\\
     State FP32 & 549 &$93.4\delta/u_\tau$ & 0.204 & 0.172 & 0.185 & 0.115\\   Convective FP16 & 550 &$84.8\delta/u_\tau$ & 0.118 & 0.103 & 0.098 & 0.111\\
     State FP16 & 683 &$42.5\delta/u_\tau$ & 3.797 & 2.594 & 3.847 & 2.073\\
     Convective E5M2 & 552 &$49.9\delta/u_\tau$ & 0.110 & 0.128 & 0.084 & 0.138\\
     Convective E4M3 & 549 &$53.1\delta/u_\tau$ & 0.229 & 0.132 & 0.210 & 0.136\\
     
     \hline
     Simson &&&&&&& \\
     Full FP64       & 546 & $43.6\delta/u_\tau$  & ---  & --- & 0.16 & 0.14 \\
     Full FP32       & 546 & $43.6\delta/u_\tau$  & 0.09 & 0.10 & 0.07 & 0.10 \\
     Convective FP16 & 544 & $43.5\delta/u_\tau$  & 0.16 & 0.18 & 0.02 & 0.12 \\
     State FP16      & 543 & $43.4\delta/u_\tau$  & 0.23 & 0.13 & 0.07 & 0.05 \\     
     Convective E5M2 & 541 & $43.3\delta/u_\tau$  & 0.35 & 0.24 & 0.20  & 0.27 \\
     State E5M2      & 337 & $26.9\delta/u_\tau$  & 20.8 & 7.5  & 20.8 & 7.6 \\     
     Convective E4M3 & 300 & $14.0\delta/u_\tau$  & 20.2 & 36.2 & 20.2 & 36.2 \\
     State E4M3      & 1284 & $7.9\delta/u_\tau$  & 15.6 & 34.5 & 15.6 & 33.9 \\
     \hline
     SSDC &&&&&&&\\
     Full FP64            & 549 & $43.95\delta/u_\tau$ & --- & ---  & 0.033 & 0.077 \\
     Combined FP32      & 550 & $43.96\delta/u_\tau$ & 0.054 & 0.082  &  0.042 & 0.113 \\
     Combined FP16      & 549 & $43.93\delta/u_\tau$ & 0.059 & 0.048  & 0.086 & 0.077\\
     Convective FP32 & 549 & $43.93\delta/u_\tau$ & 0.028 & 0.104  & 0.024 &  0.114\\
     Convective FP16 & 549 & $43.90\delta/u_\tau$ & 0.024 & 0.051  & 0.045 & 0.081\\
     \hline
    \end{tabular}
}
        \caption{Details for the different channel-flow simulations conducted in a domain of size $2\pi\delta \times 2\delta \times \pi \delta$. The reported error values (computed using Equation~\ref{eq:error_channel}) are in inner units and might be impacted by the averaging times. The compressible code \revd{(}SSDC\revd{)} is expected to have higher error levels compared to reference data~\citep{Lee_Moser_2015} due to compressibility effects. Additional tests were carried out using Simson, including the effect of Reynolds number, resolution, domain size, and time step which are shown in Table~\ref{tab:channel2}.}
    \label{tab:channel}
\end{table}

As shown in Table \ref{tab:channel}, the observations from the two sets of error measures are quite similar, hinting at the confidence in the simulations that were performed. As a side note, the slightly higher errors observed for Full FP32 in Neko compared to the cases following right after it in Table~\ref{tab:channel} could be due to the extra errors introduced when writing out the statistics in single precision ASCII in CSV format (the Full FP32 simulation was the only case where the CSV file was written out using single precision). 

\begin{figure}%[H]
    \centering
    \includegraphics[width=\linewidth]{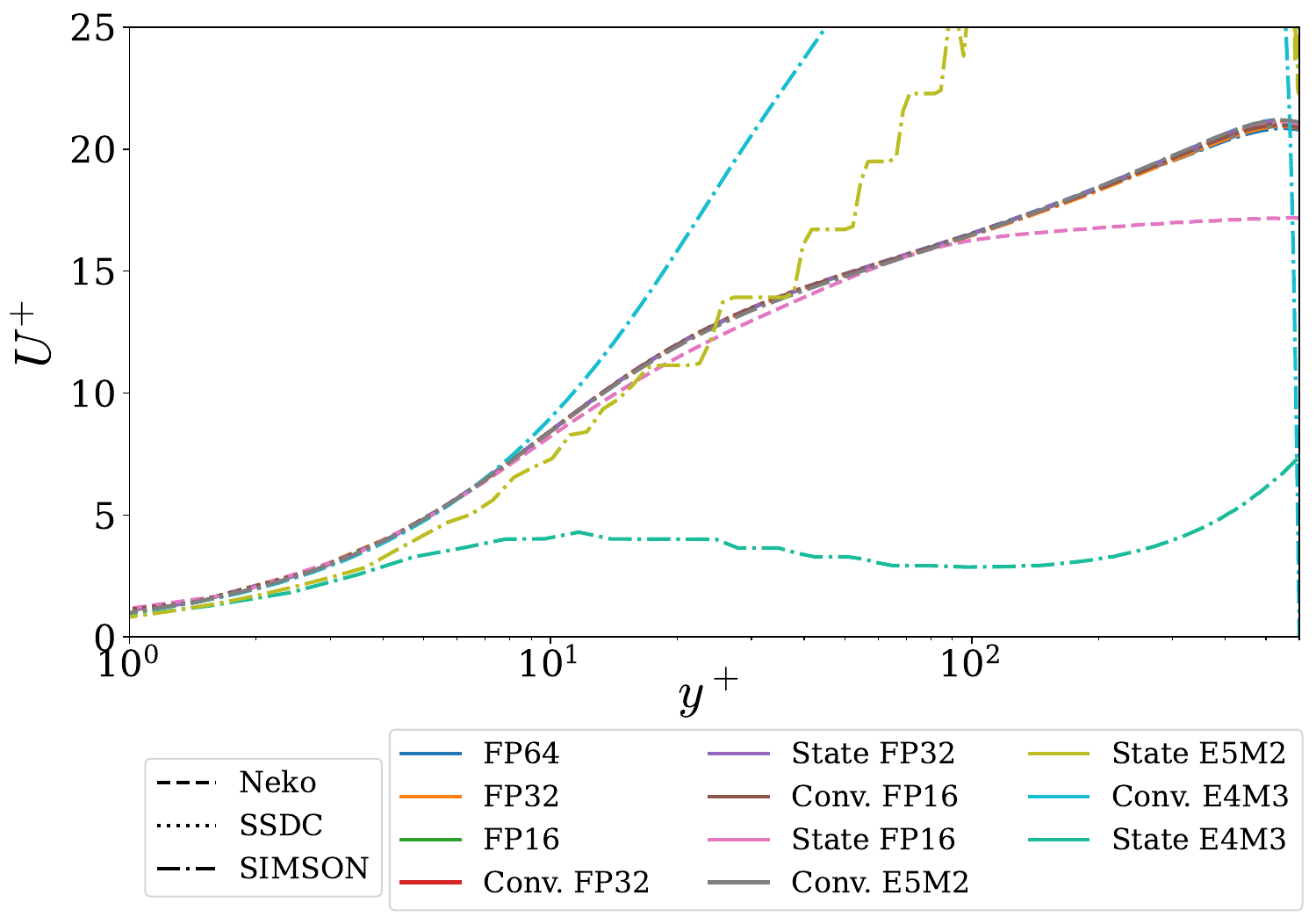} %ch550_mean.eps}
    \caption{Mean streamwise velocity profiles of the turbulent channel flow simulated using different precisions at $Re_\tau=550$ using Neko, SSDC, and Simson. All curves agree reasonably well with one another and with the data from Lee \& Moser~\citep{Lee_Moser_2015} (not shown here), except State FP16 from Neko\revdd, and State E5M2, Conv. E4M3, State E4M3 from Simson. Different codes are shown using different line styles (as indicated in the legend on the left) and different roundings are represented by different colors (as shown in the legend on the right). All these cases are also compared against each other in Table~\ref{tab:channel}.}
    \label{fig:neko_SSDC_ch550_mean}
\end{figure}

\begin{figure}%[H]
    \centering
    \includegraphics[width=\linewidth]{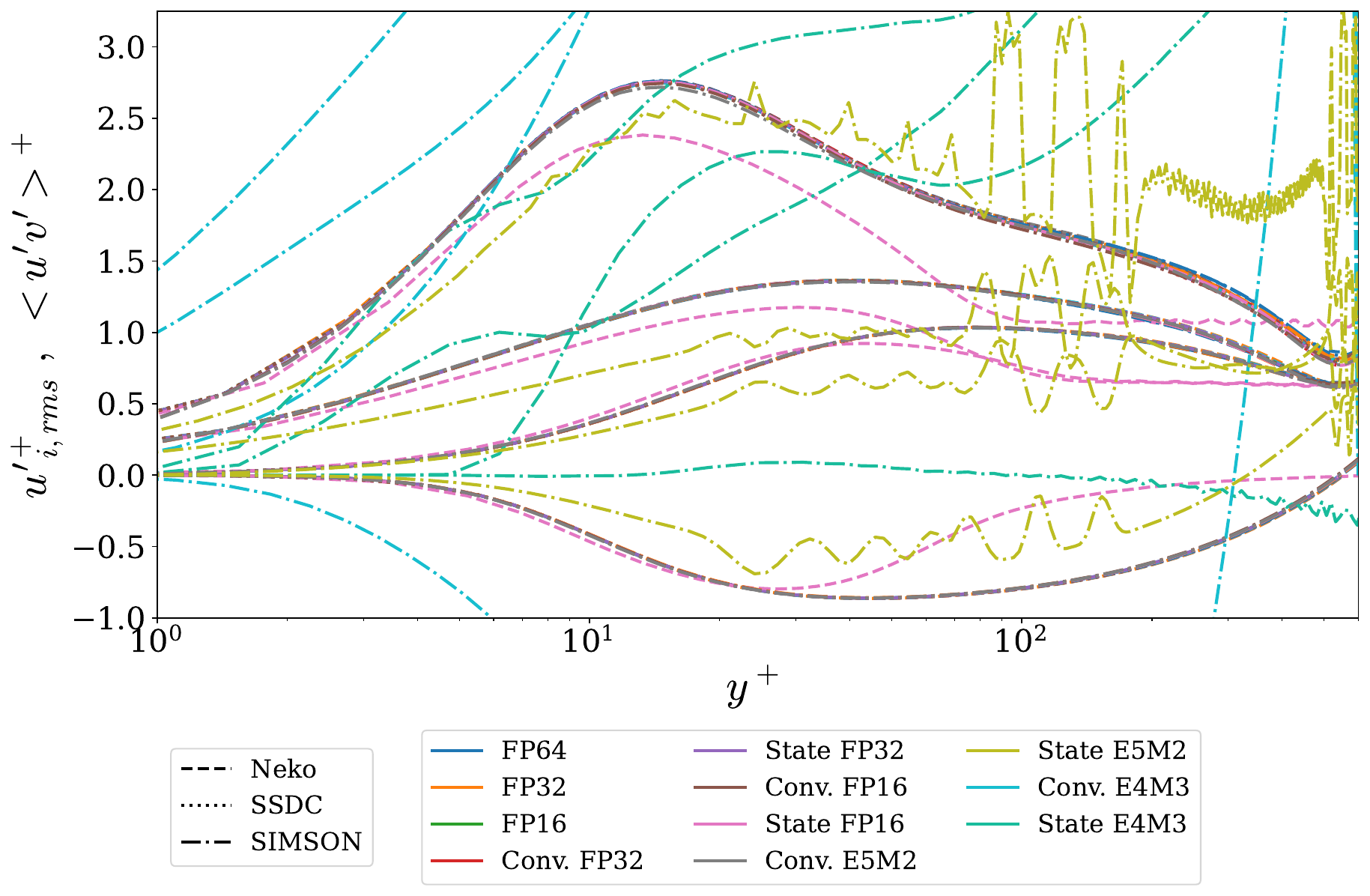}
    \caption{Root-mean-square of velocity fluctuations from the turbulent channel flow simulated using different precisions at $Re_\tau=550$ using Neko and SSDC. All curves agree reasonably well with one another and with the DNS data from Lee \& Moser~\citep{Lee_Moser_2015} (not shown here), except State FP16 from Neko, and State E5M2, Conv. E4M3, State E4M3 from Simson. Different codes are shown using different line styles (as indicated in the legend on the left) and different roundings represented by using different colors (as shown in the legend on the right). All these cases are also compared against each other in Table~\ref{tab:channel}.} 
    \label{fig:neko_SSDC_ch550_rms}
\end{figure}

Overall, the results of the different simulations were largely unaffected by low precision down to FP16. Especially for first-order moments\rev{,} the solution is not visibly sensitive. This is shown in Figure~\ref{fig:neko_SSDC_ch550_mean}  where the results from Neko, Simson\rev{,} and SSDC are compared against each other. Note that the Neko, Simson and SSDC cases that were named ``Full FPXX" or ``Combined FPXX" (as seen in Table \ref{tab:channel}) are named in Figures~\ref{fig:neko_SSDC_ch550_mean} and~\ref{fig:neko_SSDC_ch550_rms} as just ``FPXX" for the sake of brevity. % - which seemed reasonable due to lack of significant departure from expected behavior. 
This was done since they are very close in their rounded representation and indistinguishably similar in these plots. The mean streamwise velocity profiles, shown in Figure~\ref{fig:neko_SSDC_ch550_mean}, that are notably different are that of the state rounding to FP16 in Neko; and convective rounding to E4M3, state rounding to E4M3 and E5M2 in Simson. For the second-order moments shown in Figure~\ref{fig:neko_SSDC_ch550_rms}, all cases show excellent agreement with FP64 as well as with the DNS data from Lee \& Moser~\cite{Lee_Moser_2015} (not shown in the figure, but demonstrated in Table~\ref{tab:channel}), except the same cases mentioned above for the mean.% This is similar to previous observations in \citep{karp2023}. 

The lowest precision that worked fine (i.e., giving correct profiles) for the different codes are as follows: for Neko, Convective E5M2 and Convective E4M3 (but not State FP16, State E5M2, State E4M3); for Simson, Convective E5M2 (but not State E5M2, Convective E4M3 and State E4M3); for SSDC, Combined and Convective FP16 (simulations of lower precisions were not performed for turbulent channel using SSDC). 

Combining these observations, it can be concluded from the turbulent channel flow simulations that rounding the state is more sensitive than rounding the convective term alone. When rounding the state, Neko suffered more than Simson, as it produced high errors for State FP16 and diverged solutions for state roundings lower than that. In Simson, although state roundings below FP16 do not diverge, they lead to high errors. On the other hand, Neko gave accurate results for convective roundings below FP16, while only E5M2 worked well for Simson. 

It should be noted that Table~\ref{tab:channel} and Figures ~\ref{fig:neko_SSDC_ch550_mean} and ~\ref{fig:neko_SSDC_ch550_rms} show cases that did not diverge. The entire set of simulations that were carried out can be found in the Table~\ref{tab:roundings}.

\subsection{\label{sec:channel-additional} Additional assessments}

The importance of arithmetic precision was further investigated for parameters such as the Reynolds number, resolution, and domain size, and for higher-order moments and more complex statistics such as the budget terms.

The impact of the number of modes in the streamwise and spanwise directions were tested using Simson by increasing the resolution for a fixed domain size, up to a resolution of $\Delta x^+\approx9.0$ and $\Delta z^+\approx 4.5$, as well as increasing the domain size to $8\pi\delta \times 2\delta \times 3\pi\delta$ for the higher resolutions (see Table~\ref{tab:channel2}). The impact of wall resolution was also tested by decreasing the number of Chebyshev modes in the wall-normal direction (not included in Table~\ref{tab:channel2}). 
These tests were only done in FP64 and FP32 and showed no statistically significant variation in statistics such as the mean velocity or Reynolds stresses.

\begin{table}
    \centering
\resizebox{\textwidth}{!}{%
    \begin{tabular}{llllccc}
         Precision & $Re_\tau$ & Avg. time & Domain size & Resolution & 
         $\mathcal{E}_{U_{\rm LM}}$ & 
         $\mathcal{E}_{u'u'_{\rm LM}}$\\
         \hline
     Full FP64 & 544 & $145\delta/u_\tau$ & $8\pi\delta \times 2\delta \times 3\pi \delta$ & $(9.0,0.04,4.5)$ & 0.018 & 0.015 \\
     Full FP32 & 543 & $145\delta/u_\tau$ & $8\pi\delta \times 2\delta \times 3\pi \delta$ & $(9.0,0.04,4.5)$ & 0.019 & 0.013 \\
     Full FP32 & 999 & $92 \delta/u_\tau$ & $8\pi\delta \times 2\delta \times 3\pi \delta$ & $(9.8,0.03,4.6)$ & 0.10 & 0.35 \\
     Full FP32 & 998 & $18.8 \delta/u_\tau$ & $4\pi\delta \times 2\delta \times 1.5\pi \delta$  & $(9.8,0.03,4.6)$ & 0.12  & 0.11  \\
     Full FP32 & 998 & $ 51.7 \delta/u_\tau$ & $2\pi\delta \times 2\delta \times \pi \delta$ & $(9.8,0.03,4.6)$   &  0.13 & 0.11 \\
     Full FP32 & 1985 & $ 33.4\delta/u_\tau$ & $2\pi\delta \times 2\delta \times \pi \delta$ & $(10.9,0.04,6.5)$   &  0.16 & 0.09 \\
    \end{tabular}
}
        \caption{ \label{tab:channel2}
        Additional simulations of channel flow carried out using Simson.
        Resolution are reported in friction units as ($\Delta x^+, \Delta y^+_1, \Delta z^+$) for the streamwise, wall-normal (next to the wall), and spanwise resolutions.
        % Error definitions 
        }
\end{table}

The impact of the Reynolds number (scale separation) was tested by conducting a simulation using Simson compiled with FP32 at the higher Reynolds number of $Re_\tau\approx1000$ in a domain of size $8\pi\delta \times 2\delta \times 3\pi\delta$ (Table~\ref{tab:channel2}).
The overall behavior was extremely similar to $Re_\tau\approx550$ with no outstanding difference between FP32 and FP64 for the mean velocity and Reynolds stresses.

\begin{figure}
    \centering
    \includegraphics[width=0.49\linewidth,clip=true,trim=0mm 1mm 15mm 15mm]{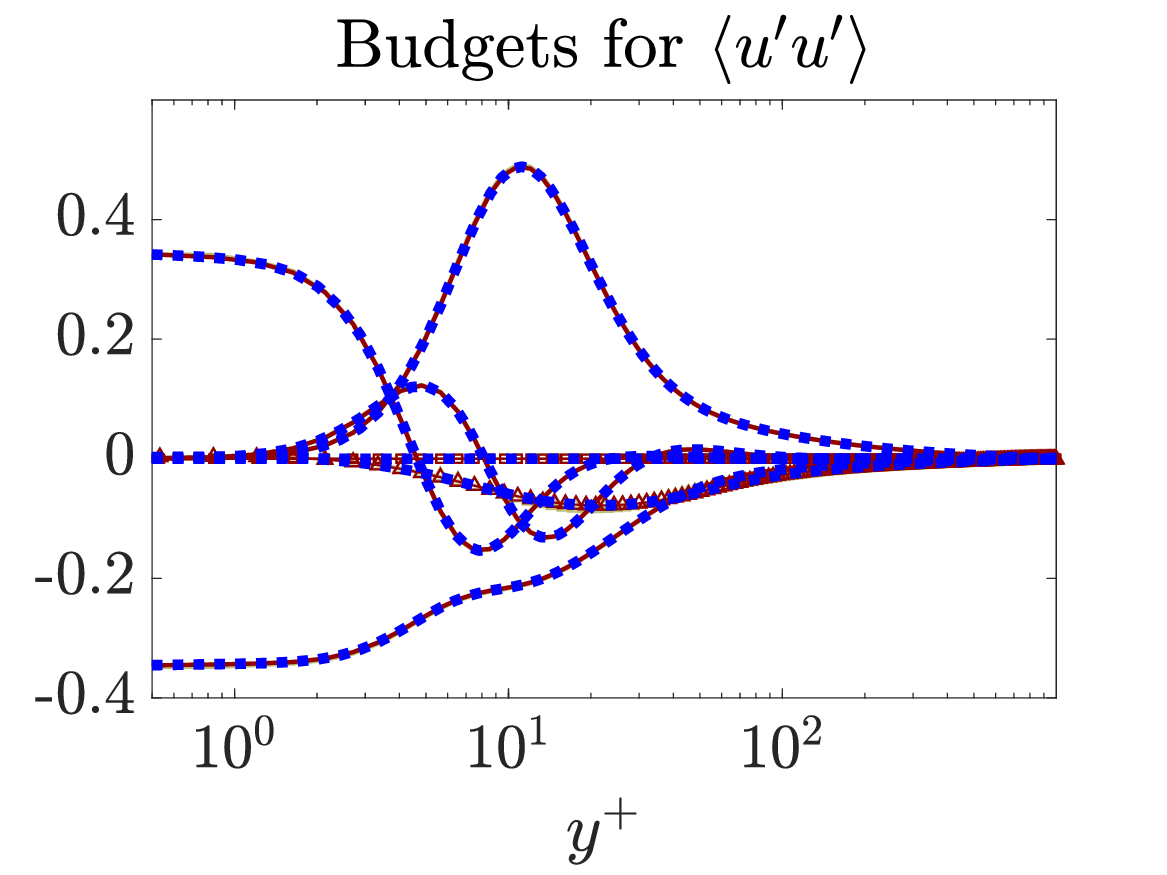}    
    \includegraphics[width=0.49\linewidth,clip=true,trim=1mm 1mm 12mm 15mm]{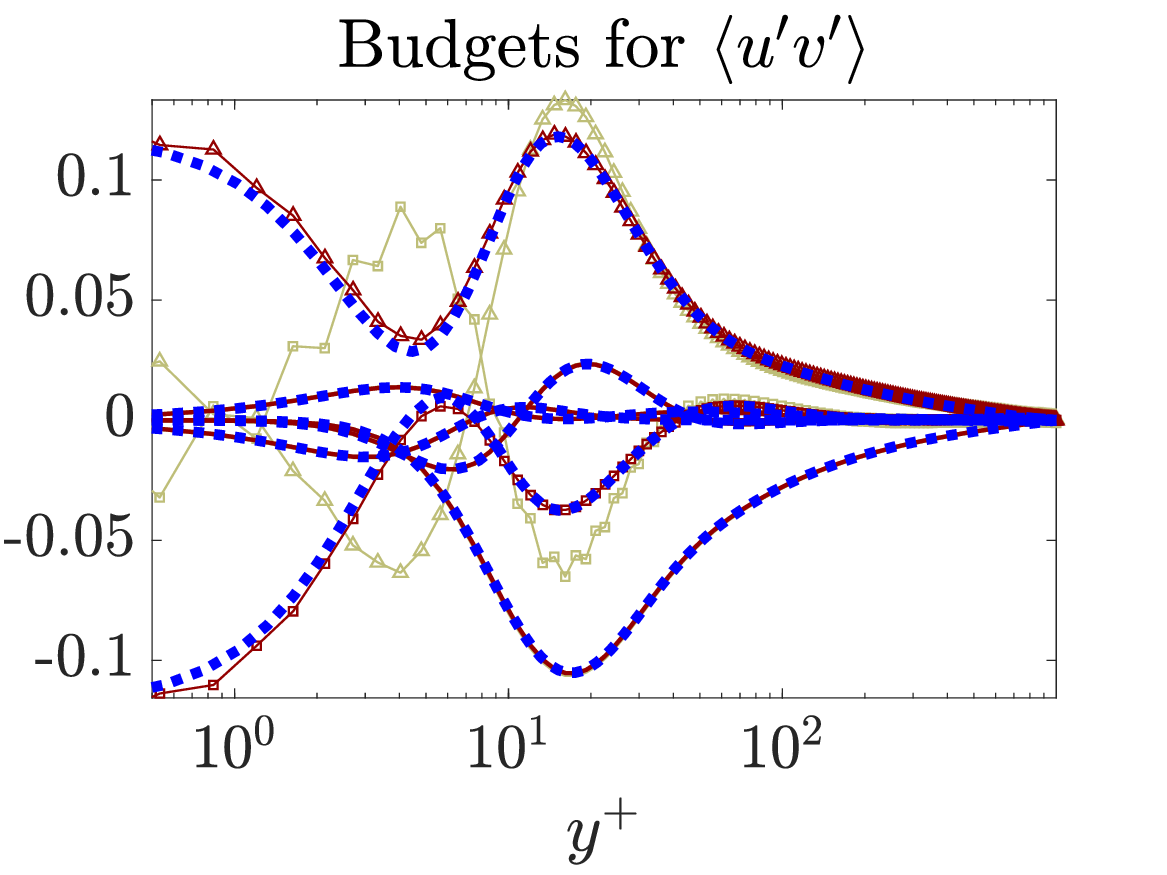}\llap{\parbox[b]{130mm}{(a)\\\rule{0ex}{-1mm}}}\llap{\parbox[b]{62mm}{{(b)}\\\rule{0ex}{-1mm}}}
    \caption{
    \label{fig:simson-channelBud}
    Budget terms in the transport equation of $\langle u'u' \rangle$ (a) and $\langle u'v' \rangle$ (b) for the
    turbulent channel flow at $Re_\tau\approx1000$ using
    full FP32 (beige) and recalculated using one time step in FP64 (dark red) compared to the reference data from Lee \& Moser~\cite{Lee_Moser_2015} (dotted blue). 
    Simulations are performed using Simson.
    Triangles and squares denote the pressure-strain and pressure transport terms, respectively. 
    }
\end{figure}

A more detailed analysis of the budget terms in the transport equation of
Reynolds stresses was also performed. 
Interestingly, the components related to pressure-velocity coupling (pressure-strain and pressure transport terms) were the only ones sensitive to the arithmetic precision at FP32, as shown in Figure~\ref{fig:simson-channelBud}. The behavior was similar in both $Re_\tau\approx550$ and $Re_\tau\approx1000$.
However, we should note that since Simson uses the velocity--vorticity formulation, the instantaneous pressure does not enter the evolution of the flow and is computed as a separate step only if needed. 
Since the budget terms related to velocity gradients were robust to precision, we hypothesized that the observed differences were likely caused by a sensitivity to precision in the Poisson solver. 
This hypothesis was confirmed by recompiling the code in FP64, restarting the $Re_\tau \approx 1000$ simulation from snapshots written in FP32, and taking one time step only to recompute the pressure in FP64.
This is shown in Fig.~\ref{fig:simson-channelBud}.
Note that there is still a small difference (around 5\%) in both pressure-related terms in the region $y^+\leq10$, which is not observed in FP64. 
Therefore, additional assessments might be necessary before adopting FP32 for producing reference DNS datasets; for example, to assess whether the correct values can be recovered by allowing a few time steps in FP64 (equivalent to switching to FP64 during runtime and before outputting the fields). 

To assess the sensitivity of the conclusions to the specific implementation in Simson, this test was repeated using Neko for the turbulent pipe flow at $Re_\tau\approx1000$ (not shown here) in a domain of length $L_z=4\pi R$ (where $R$ is the pipe radius) and resolutions of $(\Delta z^+,\Delta R^+, (R\Delta \theta)^+)\approx(5.3,0.5\rightarrow10,5.3\rightarrow10)$ in the streamwise, wall-normal, and azimuthal directions, respectively.
The choice of pipe flow was motivated by having \rev{deformed elements and thus} more complex mappings between the physical and computational space \rev{(i.e., Jacobian of the transformation to reference element is no longer diagonal and is different from point to point)}.
Interestingly, all budget terms (not shown) were extremely similar for FP32 and FP64 and matched the reference data of Yao \emph{et al.}~\cite{yao2023pipe} with no issues observed with the pressure-strain or the pressure transport term. 
\rev{Similar results were observed for smooth pipes at higher Reynolds number of $Re_\tau\approx2000$ and rough pipes with sine-shaped roughness at the wall.}
This, in fact, reinforces the previous hypothesis that the observed issues are specific to the formulation and implementation in Simson, and may not be observed, or at least can be largely avoided, in other solvers.

\begin{figure}
    \centering
    \includegraphics[width=0.49\linewidth,clip=true,trim=12mm 1mm 15mm 5mm]{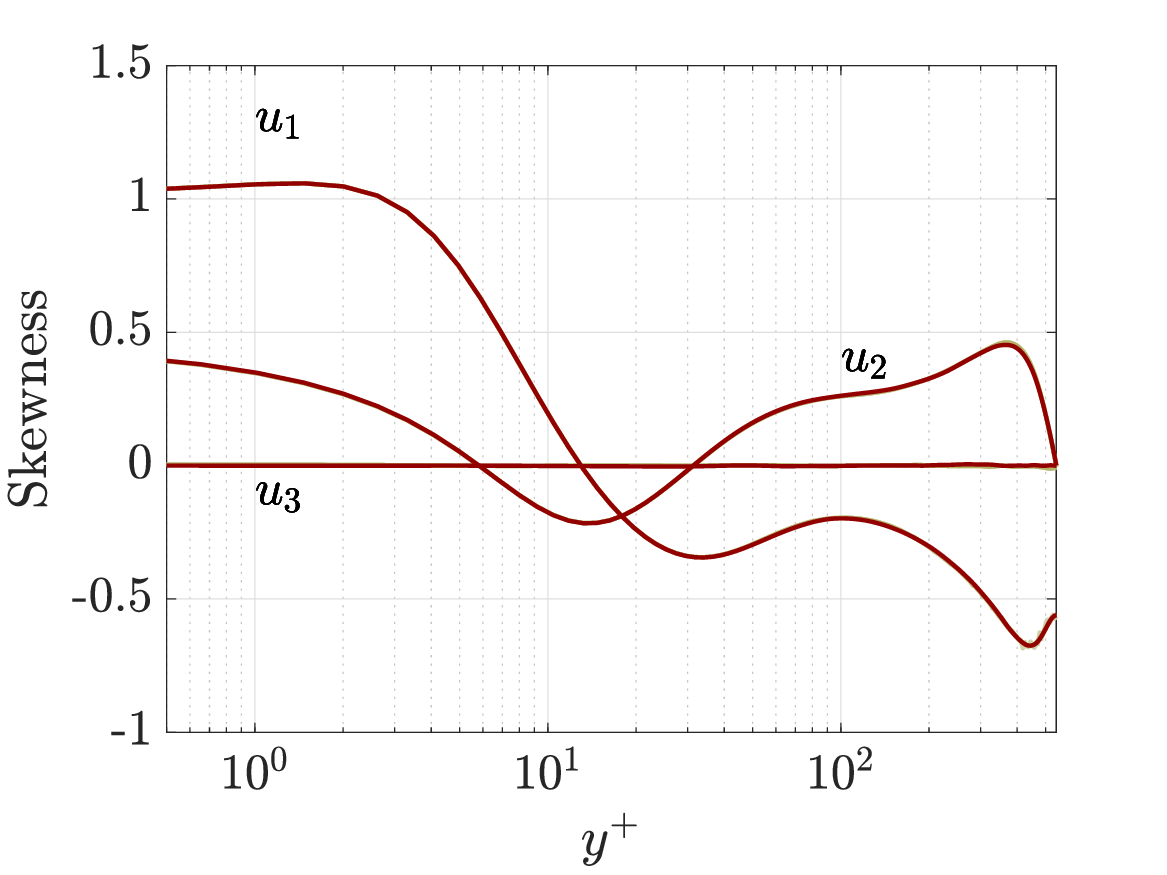}    
    \includegraphics[width=0.49\linewidth,clip=true,trim=12mm 1mm 12mm 5mm]{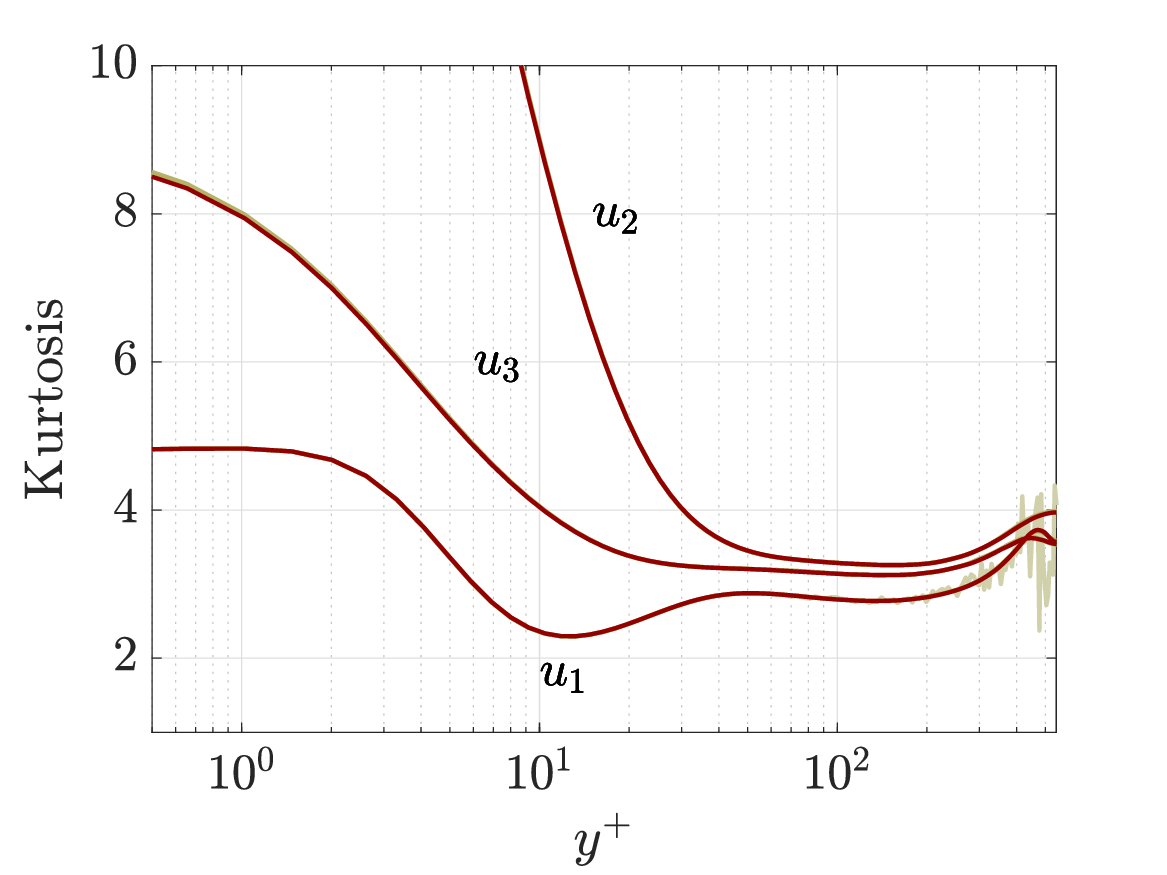}\llap{\parbox[b]{130mm}{(a)\\\rule{0ex}{-1mm}}}\llap{\parbox[b]{62mm}{{(b)}\\\rule{0ex}{-1mm}}}
    \caption{
    \label{fig:channel-moments}
    Skewness (a) and flatness (b) of velocity components in the streamwise ($u_1$), wall-normal ($u_2$), and spanwise ($u_3$) directions in the turbulent channel flow at $Re_\tau\approx550$.
    Colors from light to dark show cases that were run and post-processed in FP32, run in FP32 but restarted and post-processed in FP64,
    and run and post-processed in FP64.
    Simulations were performed using Simson with \rev{the }velocity-vorticity formulation.
    Similar behavior was observed at $Re_\tau\approx1000$, with larger fluctuations.}
\end{figure}

In addition to the budget terms, higher moments of the solution, such as the third and fourth moments, can be impacted by the precision. 
This is illustrated in Fig.~\ref{fig:channel-moments} for velocity skewness (i.e., $\langle u'^3_i\rangle / \langle u'^2_i\rangle^{3/2}$) and kurtosis (i.e., $\langle u'^4_i\rangle / \langle u'^2_i\rangle^2$), where clear fluctuations can be observed for $\langle u'^4_1\rangle / \langle u'^2_1\rangle^2$ in the region $y^+\geq100$.
This was found to be caused by two somewhat independent issues: (i) precision used for post-processing of data, and (ii) precision used for the calculation and writing of the velocity fields. 
The impact of (i) was tested by performing the entire post-processing in FP32 for a simulation done entirely in FP64, where it was observed that similar oscillations still occurred. 
Interestingly, while item (ii) seemed more serious at first, its impact could still be removed by a method similar to what was done for the budget terms, i.e., by restarting the simulation using FP64, taking one time step, and rewriting the fields. 
This procedure completely removed the oscillations, as can be observed in Fig.~\ref{fig:channel-moments}, with values that were, within statistical significance, identical to a simulation carried out and post-processed in FP64. 
These observations, combined with the absence of such oscillations from the all-normal and spanwise components of velocity (which do not have $\mathcal{O}(U_b)$ mean components, leading to $\mathcal{O}(\varepsilon U_b)$ errors), suggest that the sensitivity is most likely the result of a combination of the precision with which the velocity was written and the solver that calculates the streamwise velocity from wall-normal velocity and vorticity (as is done in the velocity-vorticity formulation).

Both the budget terms and higher velocity moments were tested for the turbulent channel flow at $Re_\tau\approx 2000$ (Table~\ref{tab:channel2}).
The observations were nearly identical to $Re_\tau\approx 550$ and $Re_\tau\approx 1000$, except for the increased sensitivity of the moments with increased Reynolds number. 
We did not test whether the correct pressure-related budget terms or the third and fourth moments could be reconstructed by one small iteration in FP64. 
We also performed a turbulent pipe flow simulation at $Re_\tau \approx 2000$ in Full FP32 using Neko.
Similarly to $Re_\tau \approx 1000$, the budget terms (not shown here) were indistinguishable from the reference data of Yao \emph{et al.}~\cite{yao2023pipe}, again hinting at the higher sensitivity of Simson to arithmetic precision. 

One should note that the results presented here do not guarantee a similar behavior for significantly higher Reynolds numbers (such as $Re_\tau\approx 10,000$ or higher), especially when generating reference quality data for the community.
However, we feel confident that for the majority of the simulations performed nowadays, e.g., for Reynolds numbers up to $Re_\tau \approx 2000$, with some minor modifications to the code, FP32 will be sufficient for the majority of the quantities of interest, up to and including turbulent stress budgets.

%%%%%%%%%%%%%%%%%%%%%%%%%%%%%%%%%%%%%%%%%%%%%%%%%%%%%%%%%%%%%%%%%%%%%

\section{Transition to turbulence} \label{sec:ts_wave}
We consider the so-called K-type transition where a laminar baseflow, as described by Schlatter et al.~ \cite{schlatter_2005, schlatter_stolz_kleiser_2006}\rev{,} is perturbed by one 2D and two oblique 3D Tollmien--Schlichting (TS) waves with amplitudes of $3\%$ and $0.1\%$ respectively (based on the laminar centerline velocity), all of which are individually stable. K-type (Klebanoff) transition refers to the instability being of fundamental type, \emph{i.e.}\ the streamwise wavenumbers of the primary and secondary instability are the same. The eigenvectors for the initial condition were computed using a Jupyter notebook \cite{schlatter_jupyter} and superposed on top of the parabolic laminar Poiseuille flow. The Reynolds number is $Re_b=3333$ based on the constant bulk velocity, which corresponds to $Re_{cl}=5000$ based on the centerline velocity $U_{cl}$ of the initial parabolic velocity profile and the channel half-height $h$. The domain size is $5.61\delta \times 2.99\delta \times 2\delta$, adjusted to fit the chosen TS waves with $\alpha_0=1.12$ and $\beta_0=2.1$ as the streamwise and spanwise fundamental wavenumbers. For the compressible codes, instead of a fixed mass flux, a constant pressure gradient forcing is applied in the streamwise direction to drive the flow. This would lead to a lower turbulent Reynolds number, but the initial growth of perturbations is only marginally affected. Thus, in all cases, a matching bulk Reynolds number of $Re_b=3333$ is maintained prior to turbulent breakdown.

\begin{figure}
    \centering
    (a)
    \includegraphics[width=0.42\linewidth]{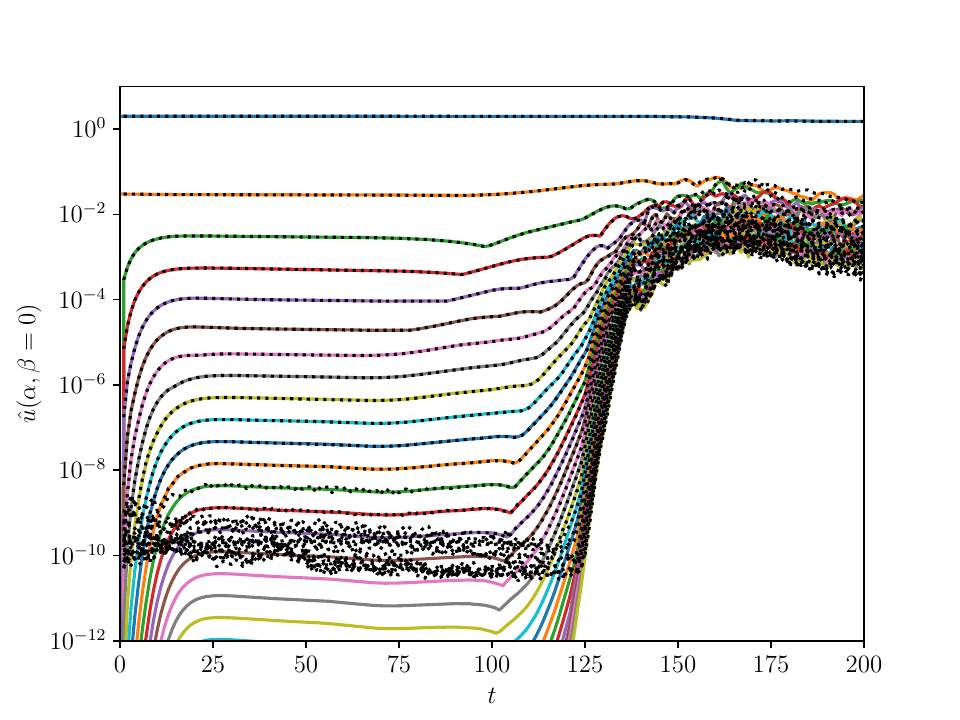}
    (b)
    \includegraphics[width=0.42\linewidth]{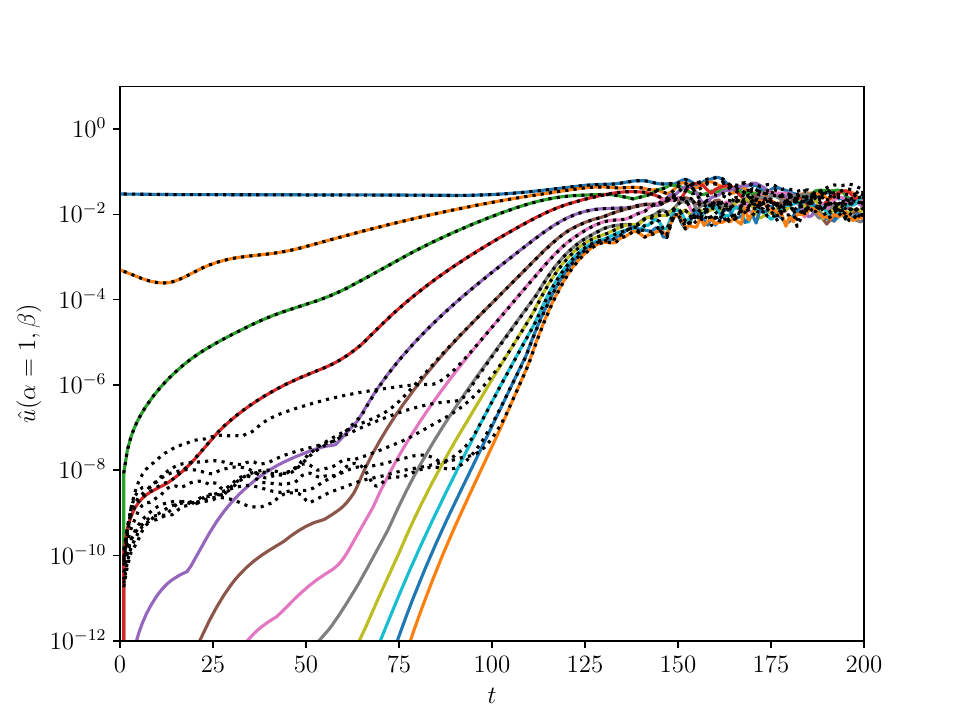}
    \caption{Evolution of Fourier modes during K-type transition, obtained with Simson. The solid lines are FP64 (double precision), and the black dotted lines \rev{are }FP32 (single precision). a) shows the two-dimensional Fourier modes  $|\hat{u}(\alpha,\beta=0)|$, starting with $\alpha = 0$ on the top. (b) Three-dimensional Fourier modes  $|\hat{u}(\alpha=1,\beta)|$ starting with $\beta=0$ on the top. The wall-normal maximum is shown in both cases.}
    \label{fig:simsonTS}
\end{figure}

We first focus on the expected behavior during transition, as illustrated in Figure \ref{fig:simsonTS}. Panel a) shows the evolution of the two-dimensional (spanwise) modes $|\hat{u}(\alpha,\beta=0)|$, for integer $\alpha=0, 1, \ldots$. It is always the maximum absolute value of the mode over the channel shown. The mean-flow modes $\alpha=0$ and $\beta=0$ are only changing at $t>150h/U_{cl}$, corresponding to the establishment of the turbulent profile with a lower centerline component. The only other non-zero mode at $t=0$ is the 2D TS wave, which has $3\%$ energy. However, due to the nonlinearity of the flow and the triadic interactions, the flow quickly establishes a saturated 2D TS wave, with higher and higher 2D modes being energized, with a weak temporal decay. 

\begin{figure}
    \centering
    \includegraphics[width=0.6\linewidth]{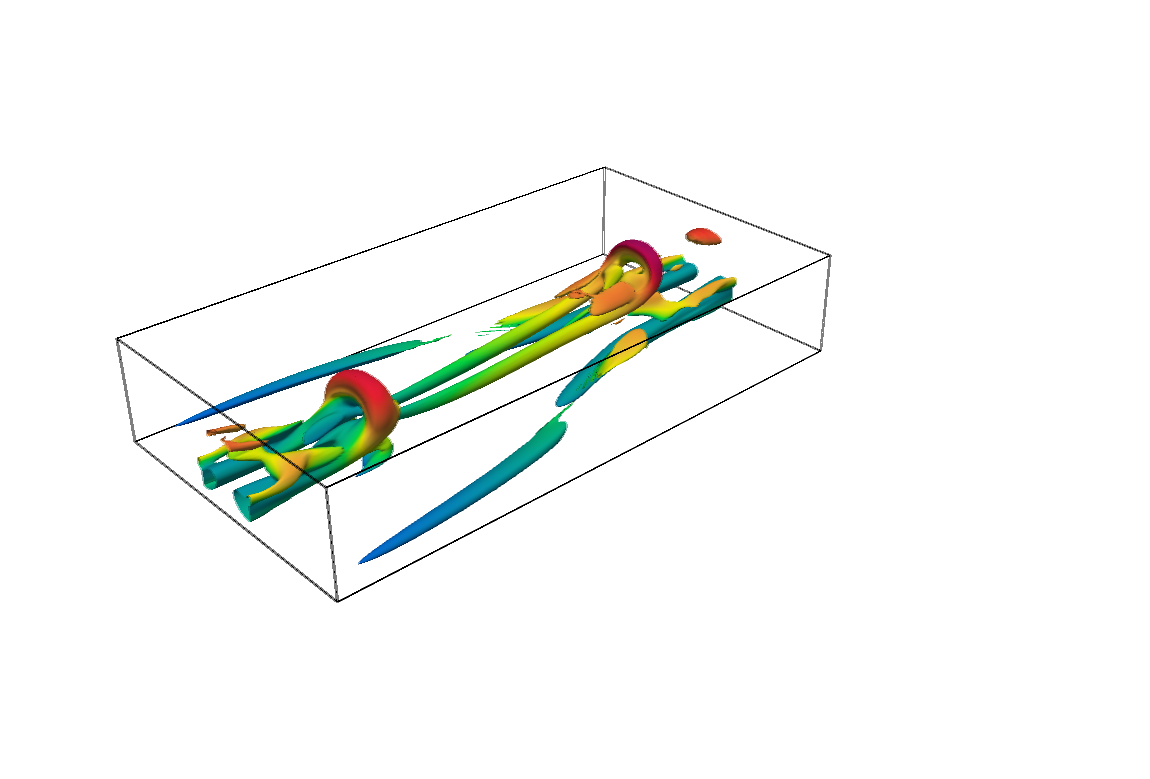}
    \caption{Three-dimensional visualization of the flow right before breakdown to turbulence ($t=136$). Iso-contours of negative $\lambda_2=-0.03$ (scaled with channel half-width and center-line velocity) colored with the streamwise velocity. Only the lower channel-half is shown. Simulation performed with Simson using FP32 precision.}
    \label{fig:TSviz}
\end{figure}

Secondary instability, initiated by the $\beta=1$ modes, leads to a quick increase in the energy in all modes ($t>120h/U_{cl}$), the formation of characterisitic hairpin vortices ($t=136 h/U_{cl}$, see Figure~\ref{fig:TSviz} )and subsequent breakdown to turbulence ($t>175 h/U_{cl}$). The double-precision arithmetic allows us to resolve numerically all modes down to machine precision ($10^{-15}$) for Simson, but saturates at around $10^{-9}$ for the other solvers. 
Reducing to single precision increases the ambient noise level to about $10^{-8}$ for Simson, and around one order of magnitude higher for the other codes. Interestingly, there seems to be no interaction between these modes that would lead to a premature growth in the physically relevant modes. In contrast, similar studies using low-resolution simulations have found a clear change in energy distribution and subsequent growth, which can be contained only using appropriate subgrid-scale models \citep{schlatter_2005,schlatter_stolz_kleiser_2006}. 
It is noteworthy to highlight that not only the Fourier amplitudes and integral quantities are seemingly not affected by the lower precision, but also the actual vortical flow structures, as shown in Fig.~\ref{fig:TSviz}. In contrast to lower-resolution simulations (as cited previously), turbulence does not appear prematurely or disrupt the flow. The sharp gradients around the hairpin heads are well resolved without artifacts. 
 
\begin{figure}
     \centering
     \includegraphics[width=1.0\linewidth]{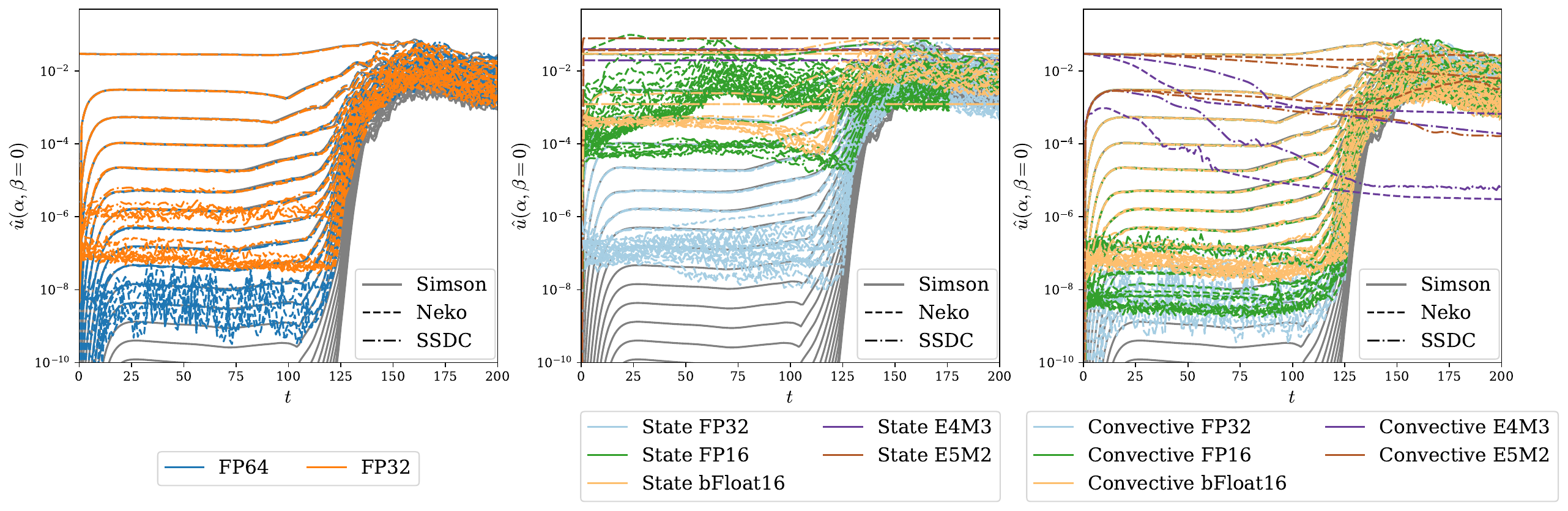}\llap{\parbox[b]{120mm}{\scriptsize{(a)}\\\rule{0ex}{7.5mm}}}\llap{\parbox[b]{81mm}{\scriptsize{(b)}\\\rule{0ex}{7.5mm}}}\llap{\parbox[b]{42mm}{\scriptsize{(c)}\\\rule{0ex}{7.5mm}}}
     \caption{Evolution in time of amplitude of 2D modes $|\hat{u}(\alpha,\beta=0)|$ for transitional case. All simulations carried out in Neko and SSDC, except the reference case in Simson, are shown in gray. Results from full FP32 and  FP64 (a), State rounding (b), and rounding of the convective term (c) are shown.}
     \label{fig:TS}
\end{figure}

From Figure~\ref{fig:TS}, we can conclude that the evolution of individual modes, as well as integral quantities such as the global friction or centerline velocities, is not dependent on the precision. For the rounding of the state and convective terms in the different solvers, we also observe that FP32 performs remarkably well, but when representing the state at lower-precision, the simulation becomes prone to stagnation (horizontal lines) or an immediate transition (State FP16). However, although the transitional case is sensitive, the amplitude of the initial conditions is still on the order of $0.1$--$1$\%, and there will be a precision-dependent limit on the smallest disturbance amplitude the simulations can capture. In addition, the geometry is  a Cartesian channel, which motivates the study of a deformed geometry, such as the separating flow around a cylinder.

\begin{figure}
    \centering
    \includegraphics[width=1.0\linewidth]{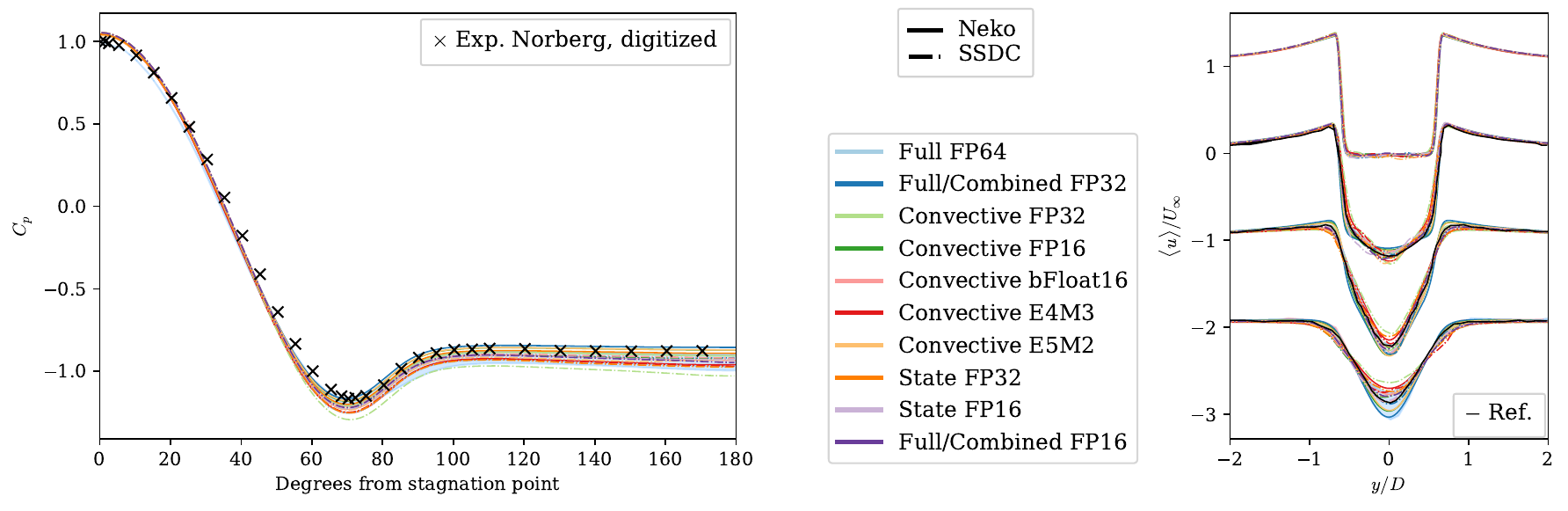}\llap{\parbox[b]{118mm}{\scriptsize{(a)}\\\rule{0ex}{0mm}}}\llap{\parbox[b]{31.5mm}{\scriptsize{(b)}\\\rule{0ex}{0mm}}}
    \caption{Profiles for the cylinder at $Re_D=3900$ in Neko and SSDC with rounding the convective term, the state, and running the entire solver in single and double precision. The $C_p$ profile in the center of the wake (a) and the wake profile at four different locations in the wake $(0.58, 1.06, 1.54, 2.02)$ (b), the blue-shaded interval is between time averages of two low-frequency modes as described by \cite{lehmkuhl2013low}.}
    \label{fig:cylinder}
\end{figure}
\begin{figure}
    \centering
    \includegraphics[width=1.0\linewidth]{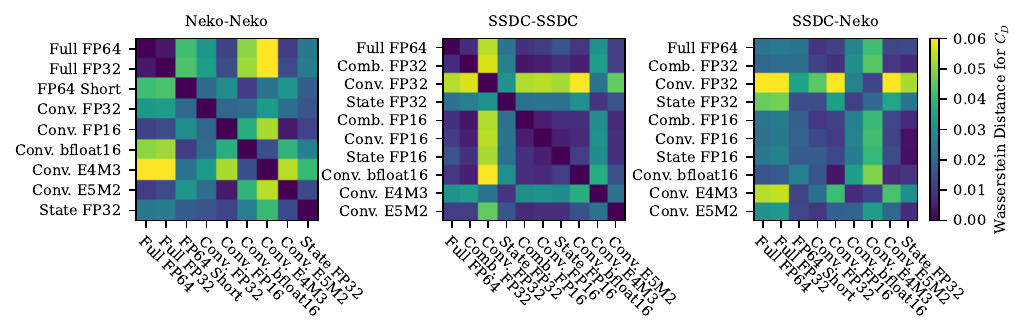}
    \caption{The Wasserstein distance between the probability distributions of the drag coefficient for the cylinder case for Neko and SSDC. A small distance means that the two distributions are similar.}
    \label{fig:wasserstein}
\end{figure}
\begin{table}
    \centering 
\resizebox{\textwidth}{!}{%
\begin{tabular}{lllllll}
     Setup & Avg. time ($D/U_\infty$) &$f_{vs}$ &$\phi_s$& $L_r$ & $\bar{C}_d$ & $-\bar{C}_{Pb}$ \\\hline
     \rev{Ref. Mode L~\mbox{\cite{lehmkuhl2013low}}} & \rev{$\sim$1166} & \rev{0.218} & \rev{87.8} & \rev{1.55} & \rev{0.979} & \rev{0.877}\\
     \rev{Ref. Mode H~\mbox{\cite{lehmkuhl2013low}}} & \rev{$\sim$1166} & \rev{0.214} & \rev{88.25} & \rev{1.26} & \rev{1.043} & \rev{0.98}\\
     \revd{\mbox{\cite{kravchenko2000numerical}}} & \revd{35} & \revd{0.21} & \revd{88} & \revd{1.35} & \revd{1.04} & \revd{0.94}\\
     \hline
Neko &&&&&&\\
Full FP64 & 300.3 & 0.2097 & 86.62 & 1.48 & 0.9926 & 0.9159 \\
Full FP32 & 319.1 & 0.2068 & 86.46 & 1.553 & 0.9911 & 0.8979 \\
Convective FP32 & 100.0 & 0.2087 & 86.97 & 1.371 & 1.025 & 0.9496 \\
Convective FP16 & 100.0 & 0.2087 & 86.62 & 1.481 & 1.004 & 0.9149 \\
Convective bFloat16 & 100.0 & 0.2087 & 87.18 & 1.271 & 1.042 & 0.9775 \\
Convective E4M3 & 73.13 & 0.204 & 87.49 & 1.182 & 1.057 & 1.003 \\
Convective E5M2 & 77.46 & 0.2054 & 86.64 & 1.472 & 1.003 & 0.9148 \\
State FP32 & 100.0 & 0.2087 & 86.8 & 1.402 & 1.016 & 0.9347 \\\hline
SSDC &&&&&& \\
Full FP64 & 300.0 & 0.2075 & 87.1 & 1.336 & 1.076 & 0.9731 \\
Combined FP32 & 100.0 & 0.2035 & 86.7 & 1.391 & 1.058 & 0.9185 \\
Combined FP16 & 100.0 & 0.2050 & 86.6 & 1.386 & 1.062 & 0.9475 \\
Convective FP32 & 100.0 & 0.2050 & 87.5 & 1.149 & 1.117 & 1.0285 \\
Convective FP16 & 100.0 & 0.2050 & 86.8 & 1.272 & 1.061 & 0.9233 \\
Convective bfloat16 & 100.0 & 0.2050 & 86.6 & 1.405 & 1.054 & 0.9202 \\
Convective E4M3 & 30.0 & 0.2099 & 87.1 & 1.205 & 1.069 & 0.9695 \\
Convective E5M2 & 100.0 & 0.2064 & 86.8 & 1.272 & 1.093 & 0.9325 \\
State FP32 & 100.0 & 0.2035 & 87.2 & 1.183 & 1.085 & 0.9756 \\
State FP16 & 100.0 & 0.2099 & 86.8 & 1.386 & 1.063 & 0.9305 \\\hline
PadeLibs &&&&&\\
Full FP64       & 69.0 & 0.2093 & 87.35 & 1.348 & 0.9932 & 0.9596\\
Convective FP16 & 36.6 & 0.2093 & 87.35 & 1.297 & 0.9932 & 0.9596\\
Convective E5M2 & 56.3 & 0.2097 & 88.86 & 1.028 & 1.073  & 1.0352\\\hline
\end{tabular}
}
   \caption{Scalar values associated with the cylinder at $Re_D=3900$. Columns correspond to each setup name, the time statistics were collected for, the separation angle $\phi_s$, the recirculation length $L_r$, the drag coefficient $\bar{C}_d$, and the base pressure coefficient $\bar{C}_{Pb}$.}
    \label{tab:cylinder}
\end{table}

%%%%%%%%%%%%%%%%%%%%%%%%%%%%%%%%%%%%%%%%%%%%%%%%%%%%%%%%%%%%%%%%%%%%%

\section{\texorpdfstring{Separated flow — Cylinder at $Re_D=3900$}{Separated flow — Cylinder at Re\_D=3900}} \label{sec:cylinder}
This section considers the flow around an infinite circular cylinder at $Re_D = {U_\infty D}/{\nu} = 3900$, where $D$ is the cylinder diameter and $U_\infty$  the free-stream velocity. We perform LES in PadeLibs, SSDC, and Neko with approximately 512 grid points along the cylinder boundary, and a spanwise length of $2\pi D$ with~$128$ grid points. There is extensive literature on this case, \rev{both experimental~\mbox{\cite{norberg1998cylinder}}, DNS~\mbox{\cite{lehmkuhl2013low}} and LES~\mbox{\cite{breuer1998cylinder,kravchenko2000numerical,witherden2020impact}}.} \revdd{showing a significant spread in the simulation results~\mbox{\citep{lehmkuhl2013low}}}{DNS results~\mbox{\cite{lehmkuhl2013low}} show a significant spread in the form of two distinct wake configurations, which requires a significant averaging time for accurate convergence.} The results are illustrated in Figure~\ref{fig:cylinder}, which shows the velocity profiles in the wake and pressure distribution on the cylinder surface. \revdd{Table \ref{tab:cylinder} compares these results with the original LES by \mbox{\cite{kravchenko2000numerical}}, and highlights}{The velocity profiles fall within the blue-shaded band defined by the two distinct solutions. In detail, Table \ref{tab:cylinder} compares the current results with the high-energy mode (Mode H), and the low-energy mode (Mode L), as described by \mbox{\cite{lehmkuhl2013low}}.} Wall quantities such as the drag coefficient and separation angle, as well as the length of the recirculation zone \rev{agree with reference data up to the mode separation issue}. Overall, the differences among the setups and solvers are comparable to the spread in the reference data. As such, for the simulations that do not diverge, \revdd{this case}{results} \revdd{indicate}{suggest} that \revdd{other}{the aforementioned} sources of uncertainty\rev{, i.e., averaging times,} are more significant than the numerical precision when it comes to \revd{LES of separating flows such as the size of the domain and, in particular, averaging times} \rev{this specific test case}. 

Isolating the impact of lower precision may become clearer with longer averaging times; however, due to the discrepancies among multiple reference data, it is unclear whether the impact of precision can be isolated. A case with a stronger consensus among the reference data and where shorter averaging times are necessary would likely be better suited to evaluate the impact of numerical precision alone.

For a more nuanced comparison, we employ the Wasserstein distance of the probability density functions of the drag coefficient $C_D$. The Wasserstein distance is a metric to compare the similarity between PDFs and has previously been employed for a similar purpose (impact of rounding and floating-point precision) in \cite{paxton2022climate} for climate models, where the similarity of the statistical description of the system is under consideration. 

By comparing the Wasserstein distance between the drag coefficient $C_D$ for the different cases, the impact of averaging times is further amplified. We show the distance between the probability distributions of $C_D$ among the different simulations in Figure \ref{fig:wasserstein}. In this plot, the variability for the simulations that were only carried out for 100 time units is clearly evident. For Neko, it is only for the longer averaging times that FP32 and FP64 consistently have a smaller Wasserstein distance than the poorer averaged results. While the convective FP32  might appear to differ, by comparing the longer FP64 simulation with exactly the same setup but averaged for a shorter time, we also observe a large difference that aligns with the low-frequency oscillations described in \cite{lehmkuhl2013low}.

%%%%%%%%%%%%%%%%%%%%%%%%%%%%%%%%%%%%%%%%%%%%%%%%%%%%%%%%%%%%%%%%%%%%%

\section{Compressible flow around a wing section} \label{sec:airfoil}
This test case considers the flow around a NACA-0012 airfoil at $Re_C = {U_\infty C}/{\nu} = 50000$, $Ma = {U_\infty}/{c_{\infty}} = 0.4$ and $\alpha = \SI{5}{\degree}$, where $C$ is the airfoil chord, $U_{\infty}$ is the free-stream velocity, $c_{\infty}$ is the speed of sound and $\alpha$ is the angle of attack.
The choice of this specific configuration is intended to analyze the effect of compressibility on reduced precision computations and to consider flows with separation, transition, and curved boundaries.

The compressible Navier--Stokes solver SSDC is considered in this test case.
Reference DNS calculations have been presented in the works of \cite{jones2008naca0012,sandberg2011naca0012}. In detail, we perform a DNS with $1080$ points on the airfoil surface and $100$ points along the spanwise direction, using a C-type grid that mimics the one from \cite{jones2008naca0012}. In particular, at the coordinate of maximum $C_f$ along the airfoil chord, $\Delta x^+=3.6$, $\Delta y^+=1.0$, and $\Delta z^+=6.0$. The grid extends in the wake direction for $5$ chord lengths and in the front with a radius of $7.3$; the spanwise dimension is $0.2$. The total number of points is roughly $\num{2.58e8}$. Given the high computational cost, only a comparison of the double-precision and single-precision computation with CPFloat is performed.

\begin{figure}[H]
    \centering
    \includegraphics[width=\linewidth]{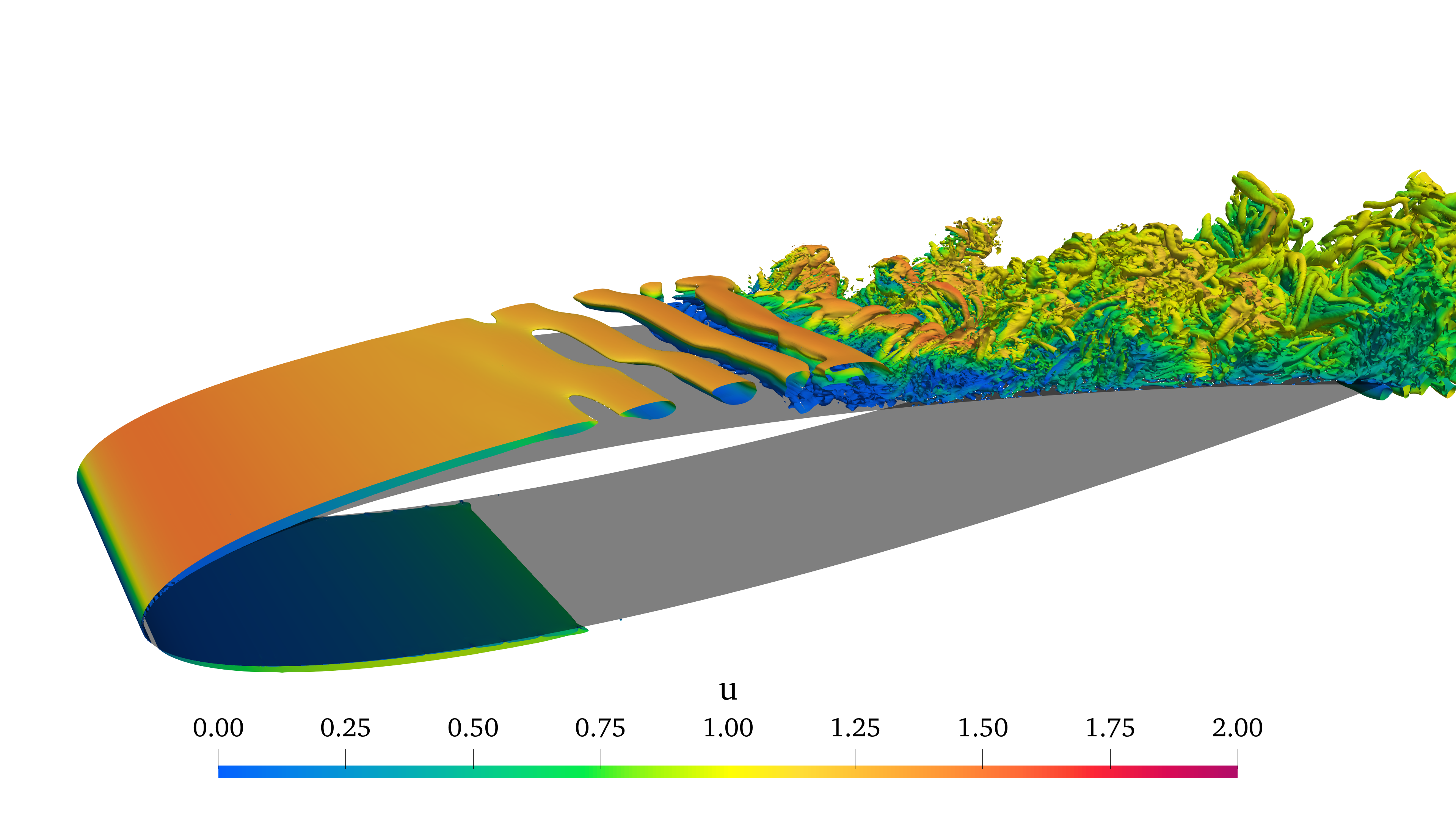}
    \caption{
    \label{fig:airfoil_qcrit}
    Laminar separation bubble with transition and turbulent reattachment observable from the iso-surfaces of the second invariant of the velocity gradient tensor ($Q = 50$) colored by streamwise velocity ($u$). 
    }
\end{figure}

This flow regime is characterized by a laminar separation bubble with transition and turbulent reattachment, as illustrated by plotting iso-contours of the second invariant of the velocity gradient tensor in Figure~\ref{fig:airfoil_qcrit}.
The flow is initialized with a preliminary two-dimensional solution. The simulation is run in FP64 for $15$ convective time units ($C/U_\infty$), after which the statistics are computed for an extra $12$ convective time units for both considered precisions.

\begin{figure}[H]
   \centering
   \subfigure[Mean pressure coefficient $\mathbf{C_p}$.]{
       \includegraphics[width=0.475\linewidth]{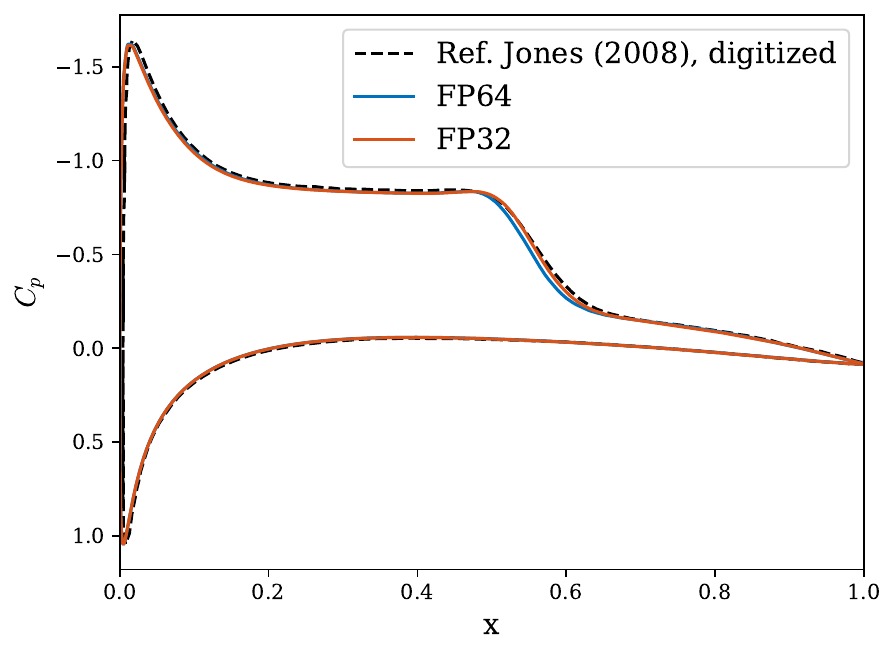}
       \label{fig:airfoil_cp}
   }
   \subfigure[Mean skin friction coefficient $\mathbf{C_f}$.]{
        \includegraphics[width=0.475\linewidth]{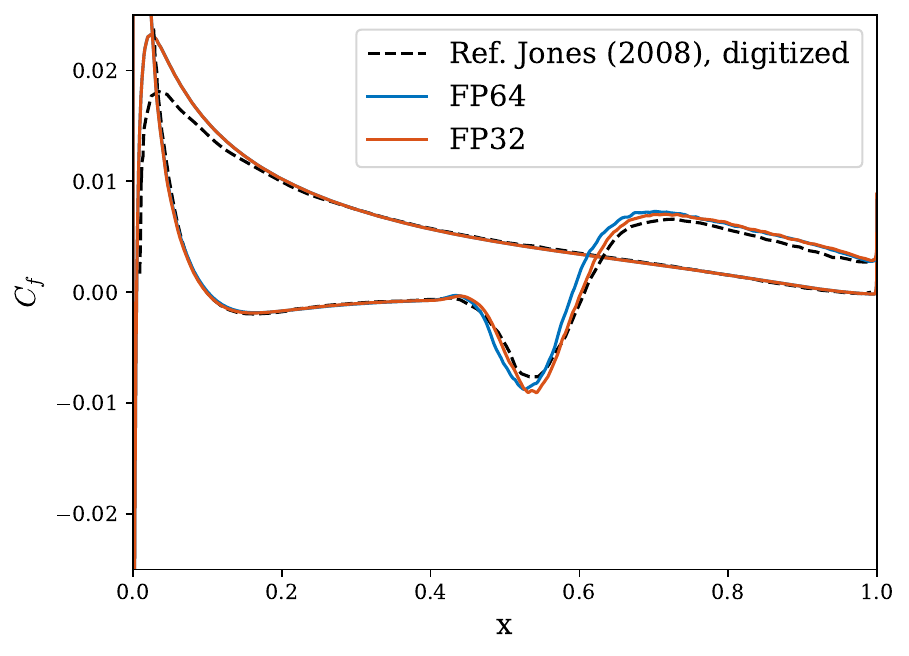}
        \label{fig:airfoil_cf}
   }
   \caption{Mean $C_p$ and $C_f$ plotted along the surface of the airfoil. Black dashed line: reference data from \cite{jones2008naca0012}. Light blue line: double precision computation, Orange line: single precision computation with PCS.}
   \label{fig:airfoil_mean_surface}
\end{figure}

\begin{figure}[H]
    \centering
    \includegraphics[width=\linewidth]{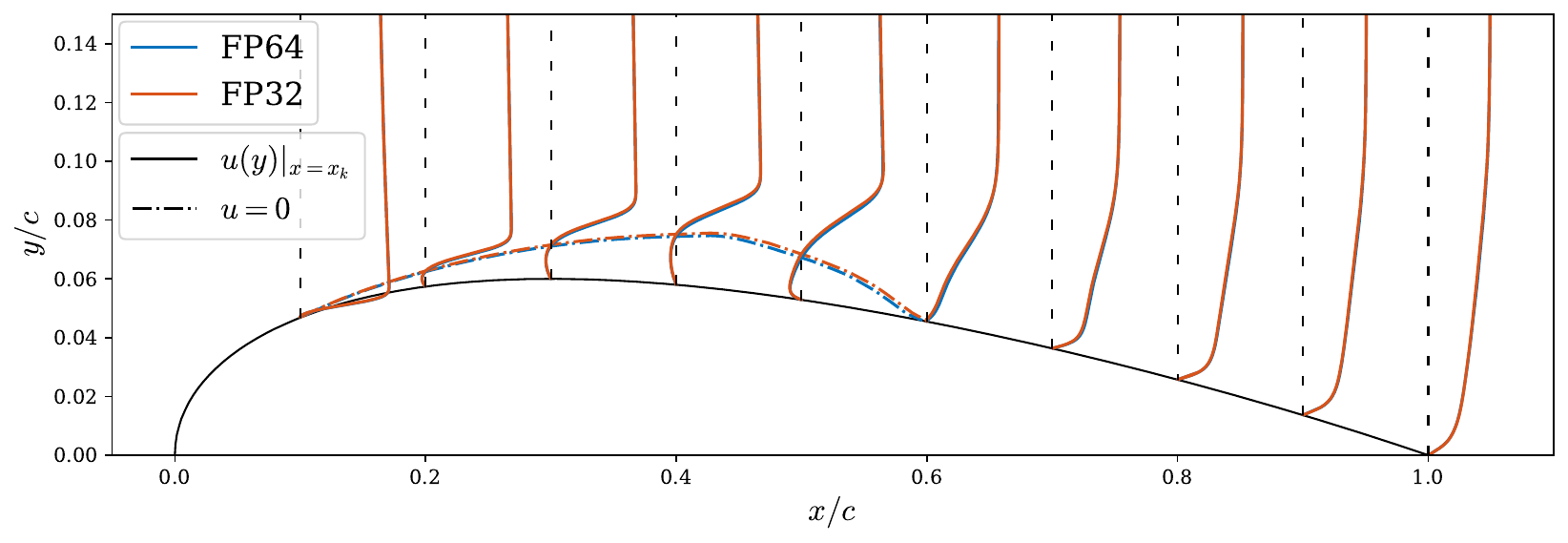}
    \caption{
    \label{fig:airfoil_velocity_profiles}
    Velocity profiles ($u(y)|_{x=x_k}$, continuous lines) along the airfoil at locations $x_k \in [0.1,1.0]$ and separation bubble ($u=0$, dot-dashed lines). FP64 in light blue, FP32 in orange.
    }
\end{figure}

\begin{figure}[H]
    \centering
    \includegraphics[width=90mm]{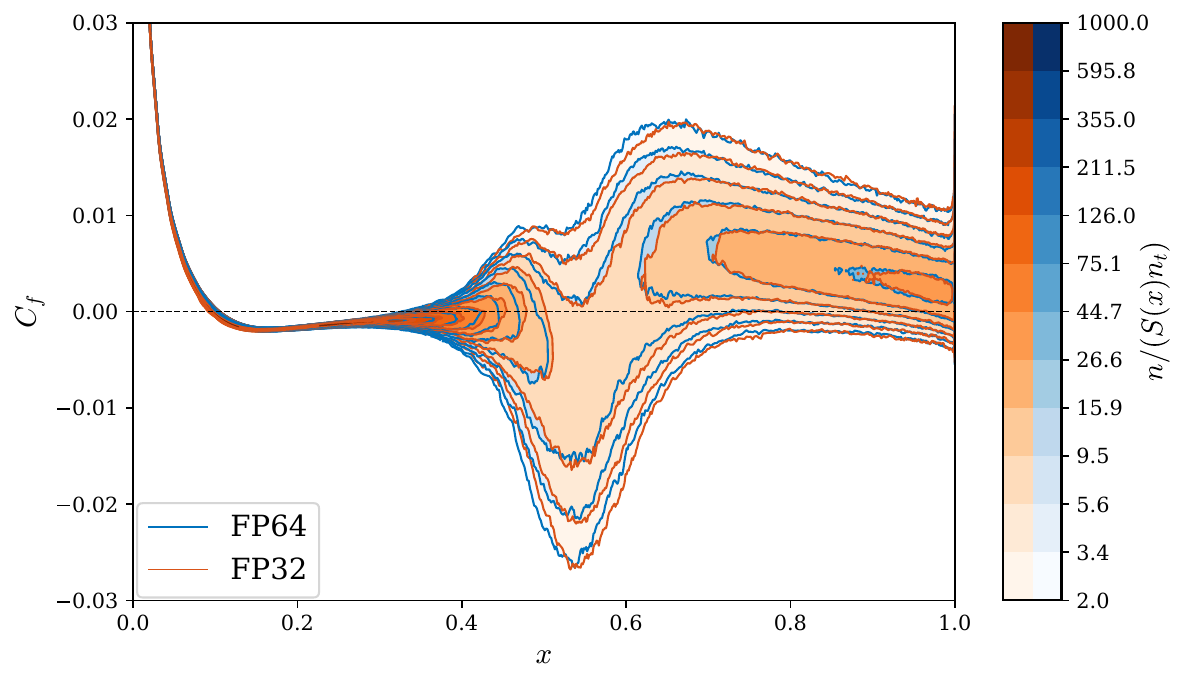}
    \caption{Contour plot of the PDF of the friction coefficient $C_f$ along the airfoil chord. The PDF is scaled, given a non-uniform bin size along $x$; more details can be found in \cite{jones2008naca0012} (Eq. 3.1, Fig. 13).}
    \label{fig:airfoil_pdf_cf}
\end{figure}

Table \ref{tab:airfoil} compares the averaged integral loads and the separation bubble extension between the two SSDC simulations and the reference data. Figure \ref{fig:airfoil_mean_surface} shows the mean pressure coefficient and skin friction coefficient on the airfoil surface. Overall, single precision computations achieve results similar to those of double precision, and both compare well to the reference. Minor differences with respect to the reference are visible in the skin friction coefficient plot, close to the leading edge on the pressure side. These can likely be attributed to a slightly different discretization of the laminar boundary layer, which is very thin in this region.
Figure \ref{fig:airfoil_velocity_profiles} shows mean velocity profiles and the separation bubble along the suction side for both precisions. Both computations achieve very similar results even for these quantities.
The time dependence of the separation is assessed considering the probability density function of the skin friction coefficient along the suction side of the airfoil, Figure~\ref{fig:airfoil_pdf_cf}. In particular, the same procedure as in \cite{jones2008naca0012}, Fig.~13c, has been applied, for example for binning. Even for this sensitive quantity, both solutions demonstrate remarkably similar results.

\begin{table}[H]
    \centering 
\begin{tabular}{llllllll}
Setup & $t_{avg} (C/U_\infty)$ & $C_L$ & $C_D$ & $C_{D_P}$ & $C_{D_{sf}}$ & $\text{x}_1|_{C_f=0}$ & $\text{x}_2|_{C_f=0}$ \\
\hline
Ref. \cite{jones2008naca0012} & 7.7 & 0.621 & 0.0358 & 0.0220 & 0.0087 & 0.0999 & 0.6066 \\
\hline
FP64 & 12 & 0.610 & 0.0355 & 0.0264 & 0.0090 & 0.1046 & 0.5965 \\
Comb.~FP32 & 12 & 0.614 & 0.0364 & 0.0275 & 0.0089 & 0.1004 & 0.6040 \\
\hline
\end{tabular}
   \caption{Scalar aerodynamic results associated with the airfoil at $Re_C=50000$ and $M=0.5$. Columns correspond to each setup name, associated time-averaging duration (in \rev{units of $C/U_\infty$}), the lift coefficient $C_L$, the drag coefficient $C_D$ (split into pressure and skin friction components), and the start and end points of the separation bubble, $\text{x}_{1,2}|_{C_f=0}$.}
    \label{tab:airfoil}
\end{table}

%%%%%%%%%%%%%%%%%%%%%%%%%%%%%%%%%%%%%%%%%%%%%%%%%%%%%%%%%%%%%%%%%%%%%

\section{Practical experiences} \label{sec:practical_experiences}

This section summarizes our experiences in running our solvers natively in FP32.
We comment on both performance gains (particularly on GPUs) and a range of issues that needed to be addressed for the solvers to run problem-free.
While the identified problems differ among the solvers, collectively they serve as useful pointers to what should be treated with extra care when implementing an FP32-capable solver.
\rev{Tools such as the Herbie project~\citep{herbie}
could be extremely valuable in catching some of these issues during code development and making the implementation more robust to reduced precisions.}

\subsection{Performance improvements with FP32 over FP64}

One of the most important practical outcomes from the simulation campaign was evaluating the performance impact of using FP32 instead of FP64 for the different test cases. Recall that FP32 was the only precision for which the rounding was available directly on the hardware and not emulated, thus for the other precisions such measurements were not possible to obtain.

We gained most experience from our simulations with Neko where a performance boost around $2\times$ was expected as most kernels operate in the memory-bound domain. However, it was observed that the performance improvement changed depending on whether server-grade or consumer GPUs were used, as illustrated in Table \ref{tab:gpus}. 

In general, it was found that a larger problem size was required to achieve a $2\times$ performance improvement on server grade GPUs when transitioning from FP64 to FP32. For the cylinder case, for example, running on the AMD Instinct MI250X GPUs, the performance was approximately $2\times$ faster with single precision. However, when a smaller turbulent channel flow case was considered, this was no longer as evident as the problem size was insufficient to hide the latency of kernel launches and oversubscribe the available computational resources. 

A comparison between the server-grade Nvidia A100 and the consumer-level Nvidia RTX 4080 revealed that, for single precision simulations, the A100 achieved a performance increase of approximately $1.5\times$ over its double precision counterpart in the smaller channel flow case. In contrast, the RTX 4080 demonstrated more than double the performance in single precision, and its FP32 performance matched that of the much more expensive A100, which operates in FP64. This suggests that it is feasible to perform DNS on consumer-grade GPUs using FP32, achieving performance comparable to that of traditional FP64 simulations on high-end server GPUs. In this case, simulations executed in FP32 on the RTX 4080 produced results equivalent to those previously obtained with FP64 on the A100---within the same runtime. 
\rev{It is worth noting that despite the higher PF32 performance ($\pi_{\rm FP32}$) of RTX4080, 
A100 is still around $2\times$ faster in FP32
due to its higher memory bandwidth.}

\begin{table}
    \centering
    \begin{tabular}{llllc}\hline
         GPU & $\beta$ & $\pi_{\text{FP}64}$  & $\pi_{\text{FP}32}$ & FP32 vs FP64\\\hline
         RTX4080&  0.72 TB/s& 0.76 TFlop/s & 48.7 TFlop/s& $2-2.5\times$\\
         A40&0.7  &0.58  & 37.4 & $2-2.5\times$\\
         A100&  1.56& 9.7  & 19.4 &$1.5-2\times$\\ \hline
    \end{tabular}
    \caption{Bandwidth to global memory (DRAM or HBM) $\beta$ in TB/s, and performance in TFlop/s }
    \label{tab:gpus}
\end{table}

\rev{The second code which was used in native FP32 compilation mode was Simson.  Note that only FP32 (and FP64) runs could be performed using hardware implementation (i.e.\ compiled for the specific precision). We nearly got the expected two-fold increase in performance, as a combination of the faster calculations and the reduced data transfer. There was an approximately 5-10\% fraction of the total execution that was not affected, including IO waiting times. The time per step with Simson, compiled and run in single precision was consistently reduced to 55\% on various CPU clusters, problem sizes and number of ranks.  
All other precisions were emulated in software which clearly increased the runtime. 
}

\subsection{Stagnation}
As previously mentioned, the issue of stagnation can become prevalent when using lower precision when summing large and small numbers.

The most straightforward and easy to address impact of lower arithmetic precision was observed, for example, in longer simulations (integration times of $\mathcal{O}(10^3)$ convective units or larger) 
with smaller time steps.
This was observed with precisions as high as FP32 and
as issues with the correct estimation of the simulation time (usually calculated as 
$t_{\rm new}=t_{\rm old}+\Delta t$)
as well as some time-dependent \texttt{if} statements
(such as those that control the calculation of additional variables at constant intervals).
While this was observed mainly in Simson on the longest simulations (integration times of $10000$ convective times or higher) when compiled with FP32, it can easily occur for other codes, especially for smaller time steps. This suggests that the integration time $t$ and its related quantities (weights etc.) should use higher precisions (FP64), even when the code is compiled with lower precision.

Another similar observation was the impact of precision on the runtime collection of solution statistics. 
This is related to the addition of the new sample to the previous set, usually done as a variant of
$S_{\rm new}=wS_{\rm old}+(1-w) s$,
where $S_{\rm new}$ and $S_{\rm old}$ are new and old statistics and $s$ is the new sample.
The weighting parameter here, $w=t/(t+\delta t)$,
is again highly impacted by precision for short sampling times $\delta t$ (i.e., frequent sampling) and long integration times $t$. 
% (with $t$ and $\delta t$ denoting time and sampling time),
This was, for instance, observed as non-zero residuals in the transport equation of Reynolds stresses
when calculated from runtime statistics, 
% when computed using statistics collected during runtime, 
while not observed when calculated from a few hundred snapshots, despite having far fewer samples.

It was also relevant when computing larger dot products in Neko, where a naive implementation can run into stagnation issues. This can be remedied through \rev{different approaches such as blocked reductions,} tree-reductions, or performing the accumulation in higher precision~\citep{blanchard2020class,higham2022mixed}. 
This again suggests that to ensure robustness and accuracy of runtime statistics, such calculations and accumulation should be performed in FP64. 

\subsection{Arithmetic errors}
\begin{figure}[H]
    \centering
    \includegraphics[width=90mm]{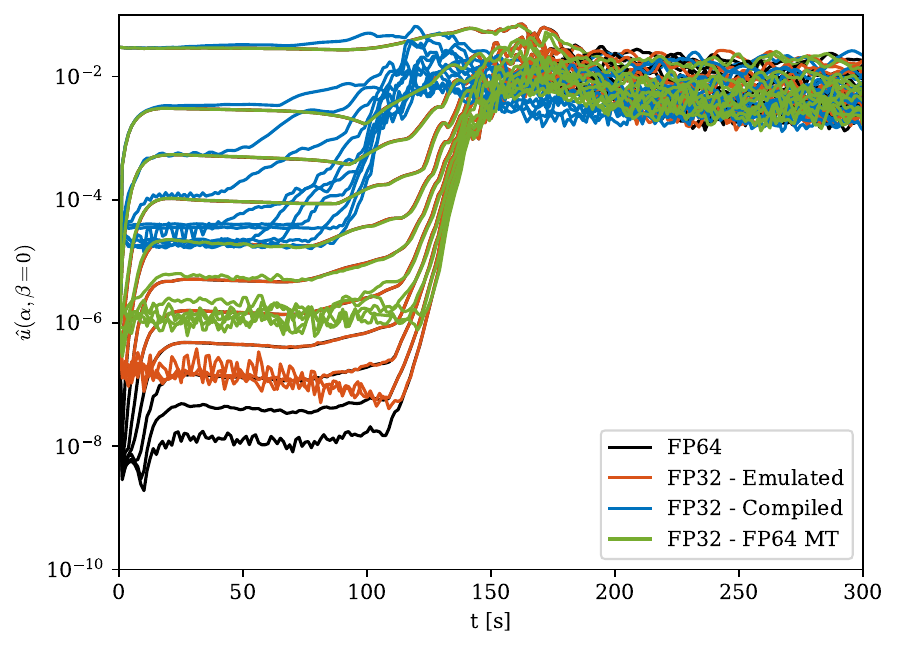}
    \caption{Evolution in time of the amplitude of 2D modes $|\hat{u}(\alpha,\beta=0)|$ for the transitional case. All simulations are carried out in SSDC. Results from full FP64 and FP32, FP32 emulated with PCS, and compiled FP32, except for the computation of metric terms (MT) in FP64, are shown.}
    \label{fig:experiences_metric_terms}
\end{figure}

Among all the computations performed with SSDC \ref{sec:ssdcintro}, the TS-wave case has shown some implementation details that can be useful in practice. The default FP64 computation is used as a reference, considering the emulated precision with the CPFloat library on both the state and operator terms, reasonable results are obtained as previously discussed. However, when compiling the code in single precision (which is not what the implementation is originally designed for), the computation exhibits nonphysical artifacts in the flow, leading to larger numerical noise and premature transition to turbulence (see Figure\ \ref{fig:experiences_metric_terms}). Upon close examination, we could trace the origin of the error to the convective term, specifically in the form of a loss of freestream preservation. In simpler terms, the computation of geometric quantities (Jacobians, cell volumes, and normals) in single precision resulted in a non-watertight grid. The correct behavior was recovered when these computations were switched back to FP64. Similar observations could also be seen when computing the Lambda-2 criterion in Neko, where a naive method for obtaining the required eigenvalues was sensitive to the lower floating point precision. The most straightforward way to remedy this was by making parts of this computation in FP64, as it was neither time nor memory critical.

For Simson, it was also evident that in the Chebyshev discretization, points are highly clustered near the ends of the interval, and the derivative matrices are fully populated. As a result, the derivative at any given point is calculated as a weighted sum of values from all other points, with both the weights and function values varying significantly across the domain.
A quick examination of the differentiation matrix for first and second derivatives using 257 points 
(the grid used for the channel flow at $Re_\tau\approx 550$)
reveals differences of approximately five orders of magnitude between terms in the same row (terms multiplied by values on wall-normal grid points), with the largest discrepancies occurring near the walls. This effect is further amplified in a velocity-vorticity formulation, where higher-order derivatives (up to the fourth order) are required.
These factors suggest a high sensitivity to numerical precision, particularly near the wall, which is likely the root cause of the observed issues with the pressure solver in Simson (Section~\ref{sec:channel-additional}). Note that the employed Chebyshev-tau method is less sensitive to poor conditioning.

Similarly, computing the streamwise and spanwise velocity components requires reconstructing values from wall-normal velocity and vorticity (with varying magnitudes) and solving a Poisson equation for the mean components. Both processes can be highly sensitive to numerical precision, which explains the sensitivity of higher moments of streamwise velocity discussed in Section~\ref{sec:channel-additional}. 

It is important to highlight that while precision was expected to influence derivatives and flow dynamics significantly, interestingly and somewhat unexpectedly, these sensitivities did not result in irrecoverable errors. 
In other words, many of the issues were found to be resolved by performing as little as a single time step in FP64.

%%%%%%%%%%%%%%%%%%%%%%%%%%%%%%%%%%%%%%%%%%%%%%%%%%%%%%%%%%%%%%%%%%%%%

\section{Conclusions}

The main outcome of this work is the demonstrated strong evidence that high-fidelity simulations of wall-bounded turbulence, including direct numerical and large-eddy  simulation (DNS and LES), do not necessarily require FP64 arithmetic and can be performed at lower precisions, such as FP32, with minimal impact on the results.
We demonstrated that this is possible not only for simpler canonical flow cases such as channels, but also for more complex flows that exhibit separation and transition. Our results are consistent across different formulations of governing equations considered in the paper, as well as underlying discretization methods and their implementations in different codes.
\revdd{Essentially perfect}{Perfect} overlap is obtained between FP64 and FP32 profiles of various quantities of interest, including, for example, turbulent kinetic energy budgets and high-order statistics.
Although we do not focus on performance analysis, we show that using FP32, significant acceleration, up to the ideal $2\times$ for a well-optimized code, is possible compared to standard FP64. In particular, in FP32, a DNS of selected cases can be run on a single consumer GPU instead of more expensive server-grade cards. 
In addition, the performance improvement is not limited to computational time but also translates to reduced storage demands.
Considering that large-scale CFD simulations utilize hundreds of millions of core hours yearly, all codes should apply significant effort to utilize lower-precision arithmetic, thus saving energy, storage, and money for the same scientific outcomes.
However, some caution must be exercised when porting existing solvers, as one likely needs to retain some of the operations in FP64 \rev{or adapt the implementation to make it more robust to arithmetic precision} (see Section~\ref{sec:practical_experiences}).

In addition, the paper explores the use of even lower precisions through software emulation.
The results are encouraging, with sometimes very low precision, like E5M2, providing good results when applied to only the convective term.
A natural way forward is to investigate the native implementation of lower precision formats  supported by modern hardware.
Success is likely conditioned on using mixed precision with the concrete realization of the latter tightly coupled with the numerical method used by the code.

On a subjective note, the results of this study stand in quite strong contradiction to our initial expectations.
We anticipated quickly finding a (high) precision threshold, after which the results become unusable.
In particular, for flows with transition, we were very skeptical about the possibility of using anything but FP64.
However, we were proved wrong, and our hope is that this paper will motivate an equally skeptical reader to give reduced precision a chance.

More generally, the question of using arithmetic precisions lower than FP64 should not be viewed as one with a binary yes or no answer, but instead as another hyper-parameter in a simulation. 
In large-scale simulations, it is often the norm to assess the influence of multiple parameters before the production simulation starts.
These parameters usually include things such as grid resolution, time-step size (\emph{i.e.}\ Courant number), domain size, residual thresholds of iterative solvers, and so on.
In our view, precision should also become one such parameter, even though the control over it is usually quite rough (i.e., a jump from FP64 to FP32 instead of a smooth transition).
In fact, one could argue that even in cases where reduced precision leads to (small) discrepancies in the results, it should still be viewed as just another source of uncertainty in the result, albeit a bias.
In that sense, viewing a computational simulation as an optimization problem to maximize the accuracy of the output for a given cost, arithmetic precision must be balanced against other sources of uncertainty such as those related to time averaging, residual, or resolution.
Similarly, in parametric studies and simulation campaigns a lower cost per simulation would enable additional simulations, which again should be balanced against other parameters for an objective function of maximizing the knowledge gained from the campaign for a given computational cost. 

Despite the broader scope of this study compared to the available literature, it is important to keep in mind that it is still fairly limited compared to the wider applications of high-fidelity simulations. 
For example, the Reynolds numbers of this study were still relatively low, and complex geometries or physics were avoided. 
Future extensions of this work could benefit from a wider variety of test cases, including supersonic and hypersonic flows, combustion, and complex geometries, among other aspects. 

The other shortcoming of this work was the lack of a quantitative and theoretical measure for the effect of precision in different regions of the domain. 
While developing such measures is relatively easy for compressible flows with explicit time stepping and no iterative solvers, it was proved difficult for an incompressible solver such as Neko with a fractional step algorithm, coarse grid solvers, and iterative methods. 
Developing a metric to quantify the impact of precision on the governing equations and the solution will be another topic for future research. 

\paragraph{Acknowledgments} This project was initiated during the CTR Summer Program 2024 in Stanford \cite{karp2024-ctr}, for which we gratefully acknowledge the financial support that enabled our participation. Computer time was provided by the National Infrastructure for Computing in Sweden (NAISS).
The authors gratefully acknowledge the HPC resources provided for this collaborative effort by the Supercomputing Laboratory at King Abdullah University of Science \& Technology (KAUST) in Thuwal, Saudi Arabia. This project has received funding from KAUST under grant No. BAS/1/1663-01-01.
The authors gratefully acknowledge the scientific support and HPC resources provided by the Erlangen National High Performance Computing Center (NHR@FAU) of the Friedrich-Alexander-Universität Erlangen-Nürnberg (FAU). The hardware is funded by the German Research Foundation (DFG). This project has received funding from the European High-Performance Computing Joint
Undertaking (JU) under grant agreement No. 101092621. The JU receives support from the
European Union’s Horizon Europe research and innovation programme and Germany, Italy,
Slovenia, Spain, Sweden and France.

\newpage

\appendix
\section{Rounding configurations}
\label{app:roundings}

Table \ref{tab:roundings} lists the rounding performed with each code for all the test cases.

    % \begin{landscape}
    \begin{table}[h!]
        \centering
        \resizebox{\textwidth}{!}{
        \begin{tabular}{|c|c|c|P{4cm}|P{4cm}|P{4cm}|P{4cm}|}
        \hline
        \textbf{CFD Code}                   & \textbf{Format}           & \textbf{Rounding type}    & \textbf{Fully developed turbulence} & \textbf{Transition to turbulence} & \textbf{Separated flow Cylinder} & \textbf{Compressible flow Airfoil} \\ \hline\hline
        \multirow{13}{*}{\textbf{Neko}}     & FP64                      & Full                      & \checkmark                          & \checkmark                        & \checkmark                       &                                    \\ \cline{3-3}
                                            & \multirow{3}{*}{FP32}     & Full                      & \checkmark                          & \checkmark                        & \checkmark                       &                                    \\
                                            &                           & Convective                & \checkmark                          & \checkmark                        & \checkmark                       &                                    \\
                                            &                           & State                     & \checkmark                          & \checkmark                        & \checkmark                       &                                    \\ \cline{3-3}
                                            & \multirow{3}{*}{FP16}     & Full                      &                                     &                                   &                                  &                                    \\
                                            &                           & Convective                & \checkmark                          & \checkmark                        & \checkmark                       &                                    \\
                                            &                           & State                     & \checkmark                          & \checkmark                        & \checkmark                       &                                    \\ \cline{3-3}
                                            & \multirow{2}{*}{bfloat16} & Convective                &                                     & \checkmark                        & \checkmark                       &                                    \\
                                            &                           & State                     &                                     & \checkmark                        & \checkmark                       &                                    \\ \cline{3-3}
                                            & \multirow{2}{*}{E4M3}     & Convective                & \checkmark                          & \checkmark                        & \checkmark                       &                                    \\
                                            &                           & State                     & \checkmark                          & \checkmark                        & \checkmark                       &                                    \\ \cline{3-3}
                                            & \multirow{2}{*}{E5M2}     & Convective                & \checkmark                          & \checkmark                        & \checkmark                       &                                    \\
                                            &                           & State                     & \checkmark                          & \checkmark                        & \checkmark                       &                                    \\\hline\hline
        \multirow{13}{*}{\textbf{PadeLibs}} & FP64                      & Full                      & \checkmark                          &                                   & \checkmark                       &                                    \\ \cline{3-3}
                                            & \multirow{3}{*}{FP32}     & Full                      &                                     &                                   &                                  &                                    \\
                                            &                           & Convective                & \checkmark                          &                                   &                                  &                                    \\
                                            &                           & State                     &                                     &                                   &                                  &                                    \\ \cline{3-3}
                                            & \multirow{3}{*}{FP16}     & Full                      &                                     &                                   &                                  &                                    \\
                                            &                           & Convective                & \checkmark                          &                                   & \checkmark                       &                                    \\
                                            &                           & State                     &                                     &                                   &                                  &                                    \\ \cline{3-3}
                                            & \multirow{2}{*}{bfloat16} & Convective                &                                     &                                   &                                  &                                    \\
                                            &                           & State                     &                                     &                                   &                                  &                                    \\ \cline{3-3}
                                            & \multirow{2}{*}{E4M3}     & Convective                &                                     &                                   &                                  &                                    \\
                                            &                           & State                     &                                     &                                   &                                  &                                    \\ \cline{3-3}
                                            & \multirow{2}{*}{E5M2}     & Convective                & \checkmark                          &                                   & \checkmark                       &                                    \\
                                            &                           & State                     &                                     &                                   &                                  &                                    \\\hline\hline
        \multirow{13}{*}{\textbf{Simson}}   & FP64                      & Full                      & \checkmark                          & \checkmark                        &                                  &                                    \\ \cline{3-3}
                                            & \multirow{3}{*}{FP32}     & Full                      & \checkmark                          & \checkmark                        &                                  &                                    \\
                                            &                           & Convective                &                                     &                                   &                                  &                                    \\
                                            &                           & State                     &                                     &                                   &                                  &                                    \\ \cline{3-3}
                                            & \multirow{3}{*}{FP16}     & Full                      &                                     &                                   &                                  &                                    \\
                                            &                           & Convective                & \checkmark                          & \checkmark                        &                                  &                                    \\
                                            &                           & State                     & \checkmark                          & \checkmark                        &                                  &                                    \\ \cline{3-3}
                                            & \multirow{2}{*}{bfloat16} & Convective                &                                     &                                   &                                  &                                    \\
                                            &                           & State                     &                                     &                                   &                                  &                                    \\ \cline{3-3}
                                            & \multirow{2}{*}{E4M3}     & Convective                & \checkmark                          & \checkmark                        &                                  &                                    \\
                                            &                           & State                     & \checkmark                          & \checkmark                        &                                  &                                    \\ \cline{3-3}
                                            & \multirow{2}{*}{E5M2}     & Convective                & \checkmark                          & \checkmark                        &                                  &                                    \\
                                            &                           & State                     & \checkmark                          & \checkmark                        &                                  &                                    \\ \hline\hline
        \multirow{18}{*}{\textbf{SSDC}}     & FP64                      & Full                      & \checkmark                          & \checkmark                        & \checkmark                       & \checkmark                         \\ \cline{3-3}
                                            & \multirow{4}{*}{FP32}     & Full                      &                                     & \checkmark                        &                                  &                                    \\
                                            &                           & Combined                  & \checkmark                          & \checkmark                        & \checkmark                       & \checkmark                         \\
                                            &                           & State                     &                                     & \checkmark                        & \checkmark                       &                                    \\
                                            &                           & Operator                  & \checkmark                          & \checkmark                        & \checkmark                       &                                    \\ \cline{3-3}
                                            & \multirow{4}{*}{FP16}     & Full                      &                                     &                                   &                                  &                                    \\
                                            &                           & Combined                  & \checkmark                          & \checkmark                        & \checkmark                       &                                    \\
                                            &                           & State                     &                                     & \checkmark                        & \checkmark                       &                                    \\
                                            &                           & Operator                  & \checkmark                          & \checkmark                        & \checkmark                       &                                    \\ \cline{3-3}
                                            & \multirow{3}{*}{bfloat16} & Combined                  &                                     & \checkmark                        & \checkmark                       &                                    \\
                                            &                           & State                     &                                     & \checkmark                        & \checkmark                       &                                    \\
                                            &                           & Operator                  &                                     & \checkmark                        & \checkmark                       &                                    \\ \cline{3-3}
                                            & \multirow{3}{*}{E4M3}     & Combined                  &                                     & \checkmark                        & \checkmark                       &                                    \\
                                            &                           & State                     &                                     & \checkmark                        & \checkmark                       &                                    \\
                                            &                           & Operator                  &                                     & \checkmark                        & \checkmark                       &                                    \\ \cline{3-3}
                                            & \multirow{3}{*}{E5M2}     & Combined                  &                                     & \checkmark                        & \checkmark                       &                                    \\
                                            &                           & State                     &                                     & \checkmark                        & \checkmark                       &                                    \\
                                            &                           & Operator                  &                                     & \checkmark                        & \checkmark                       &                                    \\ \hline
        \end{tabular}
        }
        \caption{Solvers with different roundings for each test case.}
        \label{tab:roundings}
        \end{table}
    % \end{landscape}

\end{document}